\documentclass[a4]{JHEP3}

\usepackage{amsmath}
\usepackage{amssymb}
\usepackage{axodraw}
\usepackage{epsfig}
\usepackage{mcite}

\usepackage{subfigure}

\newcommand{\be}{\begin{equation}}
\newcommand{\ee}{\end{equation}}
\newcommand{\ba}{\begin{eqnarray}}
\newcommand{\ea}{\end{eqnarray}}

\newcommand{\gev    }{\ensuremath{\mathrm{GeV}}}
\newcommand{\gevsq  }{\ensuremath{\mathrm{GeV^2}}}

\newcommand{\der}{\ensuremath{{\operatorname{d}}}}

\newcommand{\Mav}{\ensuremath{\overline{M}}}
\newcommand{\Mavav}{\ensuremath{\overline{\Mav}}}
\newcommand{\pif}{\ensuremath{(\pi^+,\pi^0,\pi^-)_f}}
\newcommand{\pii}{\ensuremath{(\pi^+,\pi^0,\pi^-)_i}}

\newcommand{\sigabs}{\ensuremath{\sigma_{\rm abs}}}
\newcommand{\ABS}{\ensuremath{\rm Abs}}

\title{Single pion electro-- and neutrinoproduction on heavy targets}

\author{
E.\ A.\ Paschos
\\
Theoretische Physik III, 
University of Dortmund, D-44221 Dortmund, Germany
\\
E-mail: \email{paschos@physik.uni-dortmund.de}
}
\author{
I.\ Schienbein
\\
Laboratoire de Physique Subatomique et de Cosmologie,
\\
Universit\'e Joseph Fourier/CNRS-IN2P3,
\\
53 Avenue des Martyrs, F-38026 Grenoble, France
\\
E-mail: \email{schien@lpsc.in2p3.fr}
}
\author{
J.-Y.\ Yu
\\
Southern Methodist University, Dallas, Texas 75275, USA
\\
E-mail: \email{yu@physics.smu.edu}
}

\abstract{
We present a calculation of single pion electroproduction
cross sections on heavy targets in the kinematic region
of the $\Delta(1232)$ resonance.
Final state interactions of the pions are taken into account
using the pion multiple scattering model of Adler, Nussinov and
Paschos (ANP model).
For electroproduction and neutral current reactions
we obtain results for carbon, oxygen, argon and iron targets and find
a significant reduction of the $W$-spectra for $\pi^0$ as compared
to the free nucleon case.
On the other hand, the charged pion spectra are only little
affected by final state interactions.
Measurements of such cross sections with the CLAS detector
at JLAB could help to improve our understanding of pion
rescattering effects and serve as important/valuable
input for calculations of single pion neutrinoproduction 
on heavy targets relevant for current and future long baseline
neutrino experiments. 
Two ratios, in Eq.\ \eqref{abs} and 
\eqref{eq:doubleratio}, will test important properties of the model.
}


\preprint{
DO-TH 07/05
\\
LPSC 07-29
\\
SMU-HEP 07-07
}

\begin{document} 

\section{Introduction}
Neutrino interactions at low and medium energies are attracting attention 
because they will be measured accurately in the new generation of experiments
\cite{Drakoulakos:2004gn,Mahn:2006ac,*Gallagher:2006ab,*Giacomelli:2006xn}.
One aim of the experiments is to measure the precise form of the cross sections and 
their dependence on the input parameters. This way we check their couplings 
and compare the functional dependence of the form factors, where deviations from 
the dipole dependence have already been established 
(see e.g.\ figure\ 1 in \cite{Paschos:2003qr} and references therein).
Deviation from the standard model predictions can arise either from properties
of the neutrinos or from new couplings of the gauge bosons to the particles 
in the target. 
Another aim of the experiments is to establish the properties of neutrinos 
including their masses, mixings 
and their fermionic nature (Dirac or Majorana particles). 
This program requires 
a good understanding of the cross sections, 
which motivated a new generation of calculations.
Since the experiments use nuclear targets, 
like ${\rm C}^{12},\, {\rm O}^{16},\, {\rm Ar}^{40},\, {\rm Fe}^{56}, ...$ it 
is necessary to understand the modifications brought about by the targets.

%The new articles concentrated on the calculation of neutrino--nucleus 
%cross sections. 
The very old calculations for quasi-elastic scattering 
and resonance excitation 
on free nucleons \cite{Rein:1981wg,Schreiner:1973mj}
have been replaced by new results where couplings 
and form factors are now better determined. For the vector couplings 
comparisons  with electroproduction data have been very useful 
\cite{Paschos:2003qr,Lalakulich:2005cs,*Sato:2003rq}. 
Axial couplings are frequently determined by PCAC.
There are already improvements and checks of the earlier quark models \cite{Lalakulich:2006sw}.
Comparisons with experimental data are also available even though the 
experimental results are not always consistent with each other 
\cite{Grabosch:1989gw,Barish:1979pj,*Radecky:1982fn,Kitagaki:1986ct}
%Alvarez-Ruso:1998hi,Leitner:2006sp}, 
but there are plans for improvements that will resolve the differences
\cite{Drakoulakos:2004gn,Mahn:2006ac}.

For reactions on nuclear targets there are modifications brought about by the 
propagation of the produced particles in the nuclear medium.
They involve absorption of particles, restrictions from Pauli blocking, 
Fermi motion and charge-exchange rescatterings. 
One group of papers uses
nuclear potentials for the propagation of the particles \cite{Alvarez-Ruso:2003gj}.
Others use a transport theory of the final particles including channels 
coupled to each other \cite{Leitner:2006sp}. 
These groups gained experience by analyzing 
reactions with electron beams (electroproduction) and adopted their methods
to neutrino reactions \cite{Leitner:2006sp}. 

Our group investigated 1-$\pi$ pion production on medium and 
heavy targets employing the pion multiple scattering model 
by Adler, Nussinov and Paschos \cite{Adler:1974qu} that was developed in order to 
understand neutral current neutrino interactions with nuclei.
This model was useful in the discovery of neutral currents and has been applied 
to predict neutrino-induced single pion production on Oxygen, Argon and Iron 
targets \cite{Paschos:2000be,Paschos:2001np,Paschos:2004qh} which are used in 
long baseline(LBL) experiments.
Among its characteristics is the importance of charge-exchange reactions 
that modify the $\pi^+:\pi^0:\pi^-$ ratios of the original neutrino-nucleon 
interaction through their scatterings within the nuclei. The presence of this 
effect has been confirmed by experiments \cite{Musset:1978gf}.
We note here that our results are 
valid for isoscalar targets. For non-isoscalar targets like lead, used 
in the OPERA experiment, it is possible to extend 
the ANP model \cite{Adler:1974wu}, 
which can be done in the future.

In this article we take an inverse route and use our calculation in 
neutrino reactions to go back to the electroproduction of pions on free 
nucleons and heavy nuclei. 
The plan of the paper is as follows.
In section \ref{sec:sec2} we summarize the neutrino production cross sections 
on free nucleons and in the $\Delta$ resonance region.
This topic has been described by several groups in the past few years. We 
present cross sections differential in several variables $E_\pi,\, Q^2$ and $W$.
We pay special attention to the spectrum $d\sigma/dE_\pi$, where we correct 
an error we found in our earlier calculation \cite{Paschos:2000be}. 
Then we obtain the 
electroproduction cross section by setting the axial coupling equal to zero and
rescaling, appropriately, the vector current contribution.

The main content of the article appears in section \ref{sec:sec3} 
%3.1 and  \ref{sec3.2} 3.2 
where we describe the salient features  
and results of the ANP model. This model has the nice property that it can be 
written in analytic form including charge exchange and absorption of pions.
This way we can trace the origin of the effects and formulate quantities which
test specific terms and parameters. 
As we mentioned above several features have been tested already, and we wish to 
use electroproduction data in order to determine the accuracy of the predictions.
We present numerical results for different target materials,
and study the quality of the averaging approximation and uncertainties of the 
ANP model due to pion absorption effects.   
We discuss how the
shape of the pion absorption cross section (per nucleon), 
an important and almost unconstrained ingredient of the ANP model,
can be delineated from a measurement of the total fraction 
of absorbed pions. 
%
%In Sec.~\ref{sec:angular} we extend our analysis to the
%case when the pion angular variables have not been integrated
%out and
%study the modification of the
%pion angular distribution caused by the intra-nuclear
%rescattering.
%%
%In Sec.~\ref{sec:reconstruction} we demonstrate that the
%ANP matrix can indeed be reconstructed from the measured
%eventrates for a heavy target and free nucleons.
%
Finally, in Sec.~\ref{sec:summary} we summarize the main
results. 
Averaged rescattering matrices 
for carbon, oxygen, argon, and iron targets
and for different amounts of pion absorption have been collected in
the appendices and are useful for simple estimates of the
rescattering effects.

\section{Free nucleon cross sections}
%\section{Cross sections for sinle pion production}
\label{sec:sec2}

%\subsection{Neutrinoproduction on free nucleons}
%\label{sec:neutrinoprod}

%\subsection{Electroproduction}
%\label{sec:electroprod}

%\subsection{Nuclear Corrections}
%\label{sec:nuccor}
In the following sections, leptonic pion production on nuclear targets 
is regraded as a two step process. In the first step, the pions are produced from 
constituent nucleons in the target with free lepton-nucleon cross sections \cite{Adler:1974qu}.
In the second step the produced pions undergo a nuclear interaction described by a 
transport matrix. Of course, the resonances themselves propagate in the nuclear medium 
before  they decay, an effect that we will investigate in the future.

The leptonic production of pions in the $\Delta$-resonance region is theoretically 
available and rather well understood as described in articles for both 
electro- and neutrino production, where 
comparisons with available data are in good agreement 
\cite{Paschos:2003qr,Lalakulich:2005cs,*Sato:2003rq,Lalakulich:2006sw,Alvarez-Ruso:1998hi,Leitner:2006sp}.
%Barish:1979pj,*Radecky:1982fn

The available data is described accurately with the proposed parameterizations. 
The vector form factors are modified dipoles \cite{Paschos:2003qr} which reproduce the helicity 
amplitudes measured in electroproduction experiments at Jefferson Laboratory \cite{Lalakulich:2006sw}.
The coupling in the axial form factors are determined by PCAC and data. Their functional 
dependence in $Q^2$ is determined by fitting 
the $\frac{\der \sigma}{\der Q^2}$ distributions.
For the vector form factors the magnetic dipole dominance for $C_3^V(q^2)$ and $C_4^V(q^2)$
gives an accurate description of the data. However, deviations with a non-zero 
$C_5^V(q^2)$ have also been established \cite{Lalakulich:2006sw}. 
This way a small (5\%) isoscalar amplitude is reproduced.

For the propose of this article we shall use a scaling relation connecting neutrino- 
to electroproduction. The weak vector current is in the same isospin multiplied with 
the electromagnetic current and the two are related as follows:
\begin{eqnarray}
<\Delta^{++}| V |p> = \sqrt{3}<\Delta^{+}|J_{em}^{I=1}|p>
= \sqrt{3}<\Delta^{0}|J_{em}^{I=1}|n> \ .  \nonumber
\end{eqnarray}
Taking into account the isospin Clebsch-Gordan factors 
for the $\Delta \rightarrow N\pi$ branchings one finds the following 
contributions of the $\Delta$-resonance to the
cross sections for $e p \to e p \pi^0$, $e p \to e n \pi^+$, 
$e n \to e p \pi^-$ and $e n \to e n \pi^0$
\begin{eqnarray}
\frac{\der \sigma^{em,I =1}}{\der Q^2 \der W} = 
\frac{8}{3}\frac{\pi^2}{G_F^2}\frac{\alpha^2}{Q^4} 
\frac{\der V^{\nu}}{\der Q^2 \der W} \times 
%\left \{ 
\begin{cases} 
\frac{2}{3} &: p \pi^0 \\
\frac{1}{3} &: n \pi^+ \\
\frac{1}{3} &: p \pi^- \\ 
\frac{2}{3} &: n \pi^0 
\end{cases} 
\label{eq:em}
\end{eqnarray}
where $\tfrac{\rm{d} V^{\nu}}{\rm{d} Q^2 \rm{d} W}$ denotes the 
cross section for the vector contribution alone to the reaction $\nu p \to \mu^- p \pi^+$.
The free nucleon cross sections in Eq.\ \eqref{eq:em} will be used in our 
numerical analysis. 
We shall call this the reduced electromagnetic formula. 
Its accuracy was tested in figure (5) of ref.\ \cite{Paschos:2003qr}.
Further comparisons can be found in \cite{Paschos:2004md}.

For studies of the pion angular distributions (or what is the same of the pion energy
spectrum in the laboratory frame) we begin with the triple differential cross section for 
neutrino production
\begin{eqnarray}
\frac{{\der} \sigma}{{\der}Q^2 {\der}W {\der}\cos{\theta}_\pi^\star}
&=& \frac{W G_F^2}{16\pi M_N^2} \sum_{i=1}^3\big(K_i \widetilde{W}_i -\frac{1}{2} 
K_i D_i (3\cos^2\theta_\pi^\star-1)\big)
\label{eq:electro}
\end{eqnarray}
with $K_i$ being kinematic factors of $W$ and $Q^2$ and the 
structure functions $\widetilde{W}_i(Q^2,W)$ and $D_i(Q^2,W)$ representing the dynamics 
for the process. All of them are found in ref.\ \cite{Schreiner:1973mj}. 
The angle $\theta_\pi^\star$ is the polar angle of the pion in the CM frame with
\begin{equation}
\cos{\theta}_\pi^\star = \frac{-\gamma E_\pi^{\rm CMS}
+E_\pi}{\beta\gamma|\vec{p}_\pi^{\ \rm CMS}|}
\end{equation}
where
\begin{equation}
|\vec{p}_\pi^{\ \rm CMS}| = \sqrt{(E_\pi^{\rm CMS} )^2 - m_\pi^2}
\quad {\text with} \quad
E_\pi^{\rm CMS} = \frac{W^2 + m_\pi^2 - M_N^2}{2 W} 
\end{equation} 
and the rest of the variables defined as 
\begin{equation}
\quad \nu = \frac{W^2 + Q^2 - M_N^2}{2 M_N}\, ,\, 
%E_\pi^{\rm CMS} = \frac{W^2 + m_\pi^2 - M_N^2}{2 W}\\
\gamma = \frac{\nu + M_N}{W}\, , \,
%\ {\text and}\  
\beta \gamma = \frac{\sqrt{\nu^2 + Q^2}}{W}.
\label{eq:gam}
\end{equation}
It is now straight-forward to convert the cross section 
differential in the solid angle to the one differential in the
laboratory energy of the pion, $E_\pi$, 
\begin{equation}
\frac{d \sigma}{d E_\pi} = \frac{1}{\gamma\beta|\vec{p}_\pi^{\ \rm CMS}|}\frac
{d \sigma}{d\cos\theta_\pi^\star}\, . 
\end{equation}
Having expressed all quantities in \eqref{eq:electro} and \eqref{eq:gam} in 
terms of $W,\, Q^2$ and $E_\pi$ it is possible to compute 
% the spectrum for 
the pion energy spectrum 
\begin{equation}
\frac{d \sigma}{d E_\pi} = 
\int_{W_{\rm min}}^{W_{\rm max}} dW \
\int _{Q^2_{\rm min}}^{Q^2_{\rm max}} dQ^2 \
\frac{d \sigma}{dQ^2 dW dE_\pi}\ \theta(phys). \label{eq:pi}
\end{equation} 
The limits of integration are given as 
\begin{eqnarray}
Q^2_{\rm min} &=& 0\, , \quad  Q^2_{\rm max} = \frac{(S-W^2)(S-M_N^2)}{S},
\nonumber\\
W_{\rm min} &=& M_N + m_\pi 
\, , \quad
W_{\rm max} \simeq 1.6\ {\rm GeV}
 \label{eq:q2}
\end{eqnarray}
where $S=M_N^2 + 2 M_N E_1$ is the center-of-mass energy squared
with $E_1$ the energy of the incoming lepton in the LAB system.
The $\theta$-function takes care of the constraints from the phase space.
We integrated the cross section for $E_\nu = 1\ {\rm GeV}$ and 
show the spectrum in figures \ref{fig:figoxygen}--\ref{fig:figiron}. 
In our earlier publication \cite{Paschos:2000be} 
the spectrum for $E_\pi$ was incorrect because 
we did not impose the phase space constraints correctly. 
The pion spectrum for charged current reactions is correctly reported 
in figure (4) in ref. \cite{Paschos:2003ej}. The discrepancy in ref. \cite{Paschos:2000be} 
has been pointed out for neutral currents in ref \cite{Leitner:2006sp}.
%

%An alternative method for obtaining the pion electron spectrum folds 
%the production of the resonance with its decay distribution into a pion and a nuclon.
%
The neutrino--nucleon and electron--nucleon cross sections will be 
used in the rest of this article in order to compute and test effects 
of nuclear corrections.
We deduce the electroproduction cross sections 
from neutrino production as in Eq.\ \eqref{eq:em}. 
For the triple differential cross section we follow the same procedure by setting 
the axial form factors to zero and using the relation
\begin{eqnarray}
\frac{\der \sigma^{em,I =1}}{\der Q^2 \der W \der E_\pi} = 
\frac{8}{3}\frac{\pi^2}{G_F^2}\frac{\alpha^2}{Q^4} 
\frac{\der V^{\nu}}{\der Q^2 \der W \der E_\pi} \times 
%\left \{ 
\begin{cases} 
\frac{2}{3} &: e p \to e p \pi^0 \\
\frac{1}{3} &: e p \to e n \pi^+ \\
\frac{1}{3} &: e n \to e p \pi^- \\ 
\frac{2}{3} &: e n \to e n \pi^0
\end{cases} 
\label{eq:em1}
\end{eqnarray}
A small isoscalar part in the electromagnetic cross section is omitted since it does not 
contribute to the $\Delta$-resonance but only to the background, which for $W<1.3\ {\rm GeV}$
is small and contributes for $1.3\ {\rm GeV}<W<1.4\ {\rm GeV}$.

%\section{...}
%\subsection{Pion rescattering in the ANP model}
%\subsection{result for various targets}
%\begin{itemize}
%\item Pion absorption
%\item pion angular distribution
%\item pion energy spectrum
%\item ....
%\end{itemize}

%\input{results_v2}
\section{Cross sections for heavy targets}
%\section{Nuclear rescattering effects}
%\section{Results for nuclear targets}
%\section{Results of the ANP model}
\label{sec:sec3}

In the following we will deal with single pion resonance 
production
in the scattering of a lepton $l$ off a 
nuclear target $T$ ($_6C^{12},\,_8O^{16},\, _{18} Ar^{40},\, _{26} Fe^{56}$), i.e., 
with the reactions 
\begin{equation}
l +T\rightarrow l' +T^\prime +\pi^{\pm,0}
\label{eq:reaction}
\end{equation}
where $l'$ is the outgoing lepton and $T^\prime$ a final nuclear state.
Furthermore, in our analysis of nuclear rescattering effects 
we will restrict
ourselves to the region of the $\Delta(1232)$ resonance, 
$1.1\ \gev < W < 1.4\ \gev$, 
and to isoscalar targets with equal number of protons and neutrons.

\subsection{Pion rescattering in the ANP model}
% Matrix relation
According to the ANP model \cite{Adler:1974qu,Schienbein:2003xy}
the final cross sections for pions $\pif$ can be related 
to the initial cross sections 
$\pii$ for a {\em free nucleon} target in the simple form
\begin{equation}
\left(\begin{array}{c}\displaystyle
{\der \sigma(_ZT^A;{\pi^+})\over \der Q^2\der W}\\
\displaystyle{\der \sigma(_ZT^A;{\pi^0})\over 
\der Q^2\der W}\\
\displaystyle{\der \sigma(_ZT^A;{\pi^-})\over 
\der Q^2\der W}
\end{array}\right)_{\rm f}
%={\underbrace{\betont M}_{\rm\zitat ANP model} 
= M[T; Q^2,W]\ 
\left(\begin{array}{c}\displaystyle
{\der \sigma(N_T;{\pi^+})\over \der Q^2\der W}\\
\displaystyle{\der \sigma(N_T;{\pi^0})\over \der Q^2\der W}\\
\displaystyle{\der \sigma(N_T;{\pi^-})\over \der Q^2\der W}
\end{array}\right)_{\rm i}
\label{eq:fac}
\end{equation}
with 
\begin{equation}
{\der \sigma({N_T};\pm 0)\over \der Q^2\der W} 
= {{Z}{\der \sigma({p};\pm 0)
\over \der Q^2\der W}
+ {(A-Z)}{\der \sigma({n};\pm 0)
\over \der Q^2\der W}} 
\label{eq:free}
\end{equation}
where the free nucleon cross sections are averaged over the
Fermi momentum of the nucleons.\footnote{However, the Fermi motion
has a very small effect on the $W$ distribution and we neglect it
in our numerical analysis.
On the other hand,
effects of the Pauli exclusion principle have been absorbed into
the matrix $M$ and are taken into account.}
For an isoscalar target
the matrix $M$ is described by three independent parameters
$A_p$, $d$, and $c$
in the following form \cite{Adler:1974qu}
\begin{equation}
M = A_p
\left(\begin{array}{ccc}
1-c-d & d & c 
\\
d & 1-2 d & d
\\
c & d & 1-c-d
\end{array}\right)\ , \label{eq:M}
\end{equation}
where $A_p(Q^2,W) = g(Q^2,W)\times f(1,W)$.
Here, $g(Q^2,W)$ is the Pauli suppression factor 
and $f(1,W)$ is a transport function for equal populations
of $\pi^+, \pi^0, \pi^-$ which depends on the absorption
cross section of pions in the nucleus.
The parameters $c$ and $d$ describe the charge exchange contribution.
The final yields of $\pi$'s depend on the target material and the 
final state kinematic variables, i.e., $M = M[T; Q^2,W]$.

%\subsection{Averaging Approximation}
% Averaging Approximation
In order to simplify the problem it is helpful to 
integrate the doubly differential cross sections of Eq.\ \eqref{eq:fac}
over $W$ in the $(3,3)$ resonance region, say, 
$m_p + m_\pi \le W \le 1.4\ \gev$.
In this case Eq.\ \eqref{eq:fac} can be replaced by an equation of
identical form
\begin{equation}
{{\left(\begin{array}{c}\displaystyle
{\der \sigma(_ZT^A;{\pi^+})\over \der Q^2}\\
\displaystyle{\der \sigma(_ZT^A;{\pi^0})\over 
\der Q^2}\\
\displaystyle{\der \sigma(_ZT^A;{\pi^-})\over 
\der Q^2}
\end{array}\right)}_{\rm f}} 
%={\underbrace{\betont M}_{\rm\zitat ANP model} 
= \Mav[T; Q^2]\ 
{{\left(\begin{array}{c}\displaystyle
{\der \sigma(N_T;{\pi^+})\over \der Q^2}\\
\displaystyle{\der \sigma(N_T;{\pi^0})\over \der Q^2}\\
\displaystyle{\der \sigma(N_T;{\pi^-})\over \der Q^2}
\end{array}\right)}_{\rm i}}
\label{eq:appx1}
\end{equation}
%\TODO{NOT really needed. Remove ???}
where the matrix $\Mav[T; Q^2]$ can be obtained by averaging
the matrix $M[T; Q^2,W]$ over $W$ with the leading $W$-dependence coming from
the $\Delta$ resonance contribution. 
%to the doubly differential 
%cross section \cite{Adler:1974qu}. 
Moreover,
we expect the matrix $M$ to be a slowly varying function
of $Q^2$ (for $Q^2 \gtrsim 0.3\ \gevsq$). For this reason
we introduce a  
second  averaging over $Q^2$ and define 
the double averaged matrix $\Mavav[T]$
which is particularly useful for giving a simple description
of charge exchange effects in different nuclear targets.
In the double-averaging approximation (AV2) the final cross sections 
including nuclear corrections are expressed as follows:
%calculated in the following way:
\begin{equation}
{{\left(\begin{array}{c}\displaystyle
{\der \sigma(_ZT^A;{\pi^+})\over \der Q^2 \der W}\\
\displaystyle{\der \sigma(_ZT^A;{\pi^0})\over 
\der Q^2 \der W}\\
\displaystyle{\der \sigma(_ZT^A;{\pi^-})\over 
\der Q^2 \der W}
\end{array}\right)}_{\rm f}} 
%={\underbrace{\betont M}_{\rm\zitat ANP model} 
= \Mavav[T]\ 
{{\left(\begin{array}{c}\displaystyle
{\der \sigma(N_T;{\pi^+})\over \der Q^2 \der W}\\
\displaystyle{\der \sigma(N_T;{\pi^0})\over \der Q^2 \der W}\\
\displaystyle{\der \sigma(N_T;{\pi^-})\over \der Q^2 \der W}
\end{array}\right)}_{\rm i}}\, .
\label{eq:av2}
\end{equation}
We note that the cross sections are differential in two variables 
while the matrix $\Mavav[T]$ is the average over these variables.

% ANP model
The above discussion will be 
%is quite general and can be 
used for a phenomenological description of nuclear rescattering
effects.
On the other hand, in Ref.\ \cite{Adler:1974qu} a dynamical
model has been developed to calculate the charge exchange matrix
$M$. As an example, 
for oxygen the resulting matrix in the double-averaging
approximation is given by
\begin{equation}
\Mavav(_8O^{16}) = \overline{\overline{A_p}}\left(\begin{array}{ccc}
{0.788} & {0.158} & {0.0537} \\
{0.158} & {0.684} & {0.158} \\
{0.0537} & {0.158} & {0.788}
\end{array}\right).
\label{eq:M1}
\end{equation}
with $\overline{\overline{A_p}}=0.766$, which contains the averaged
Pauli suppression factor and absorption of pions in the nucleus.
There are various absorption models described in the original article. 
Two of them are distinguished by the energy dependence of the absorption 
cross section beyond the $\Delta$ region. 
In model [A] the absorption increases as $W$ increases while in [B] 
it decreases for large $W$'s (beyond the $\Delta$ region). 
%
%by renormalizing absorption model (B) \cite{Silbar:1973em} 
%by a factor $\simeq 0.3$ \cite{Schienbein:2003xy}.
%Note that the absorption model A in 
%Ref.\ \cite{Sternheim:1972ad} 
%increases linearly beyond the $\Delta$ region, while the contribution of 
%absorption model B decreases.
A comparison of the two absorption models (A) and (B) can be found in
\cite{Schienbein:2003xy}. 
Since the fraction of absorbed pions is still rather uncertain we
provide in the appendices ANP matrices for different amounts of 
absorption.
These matrices are useful to 
obtain an uncertainty band for the expected nuclear corrections.

\subsection{Results for various targets}
In this section we present numerical results for 
1-pion leptoproduction differential cross sections 
including nuclear corrections 
using the ANP model outlined in the preceding section.

\subsubsection{Neutrinoproduction}
\begin{figure}[h]
\centering
\vspace*{-0.1cm}
\includegraphics[angle=0,width=8.9cm]{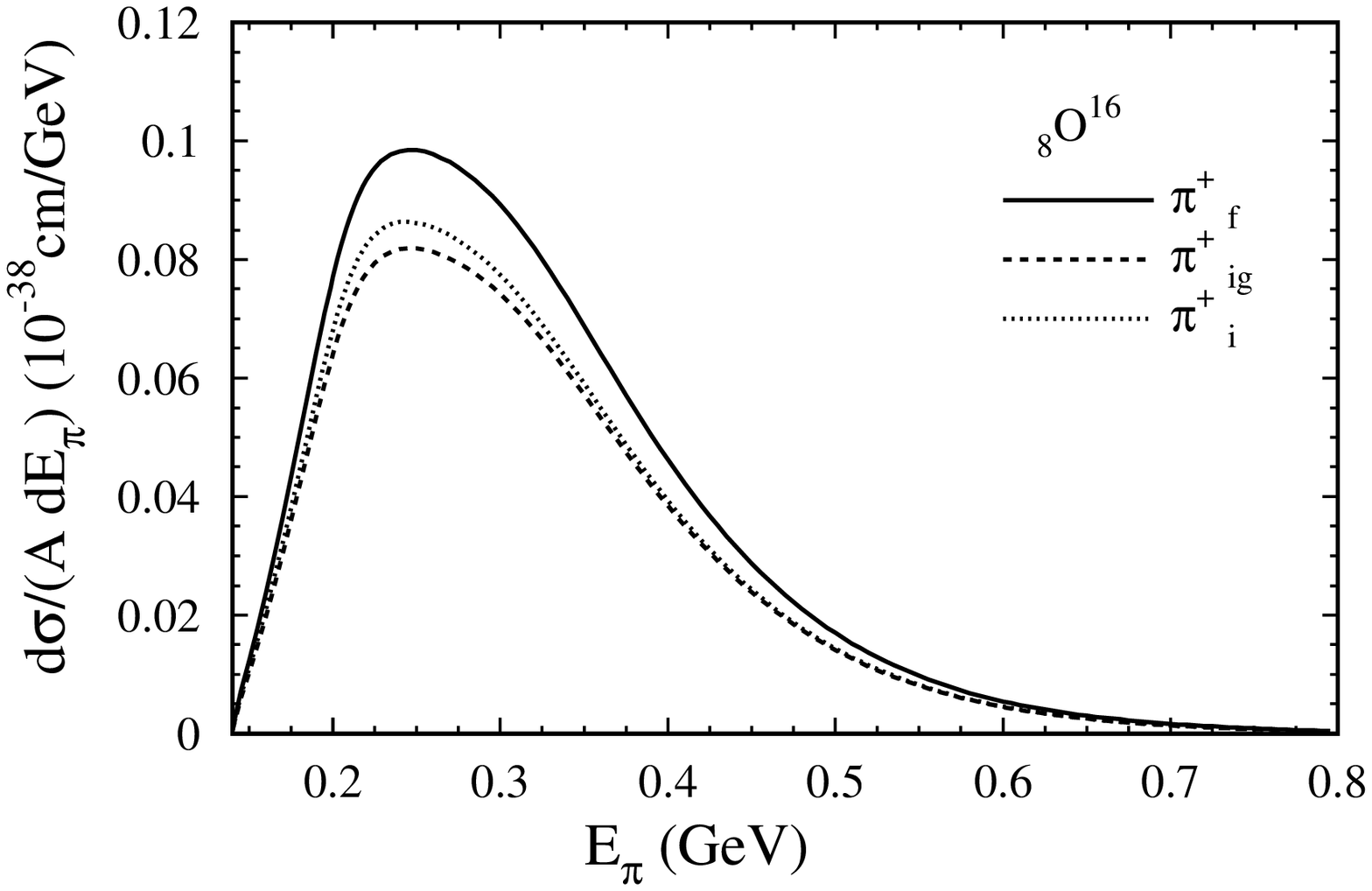}
\includegraphics[angle=0,width=8.9cm]{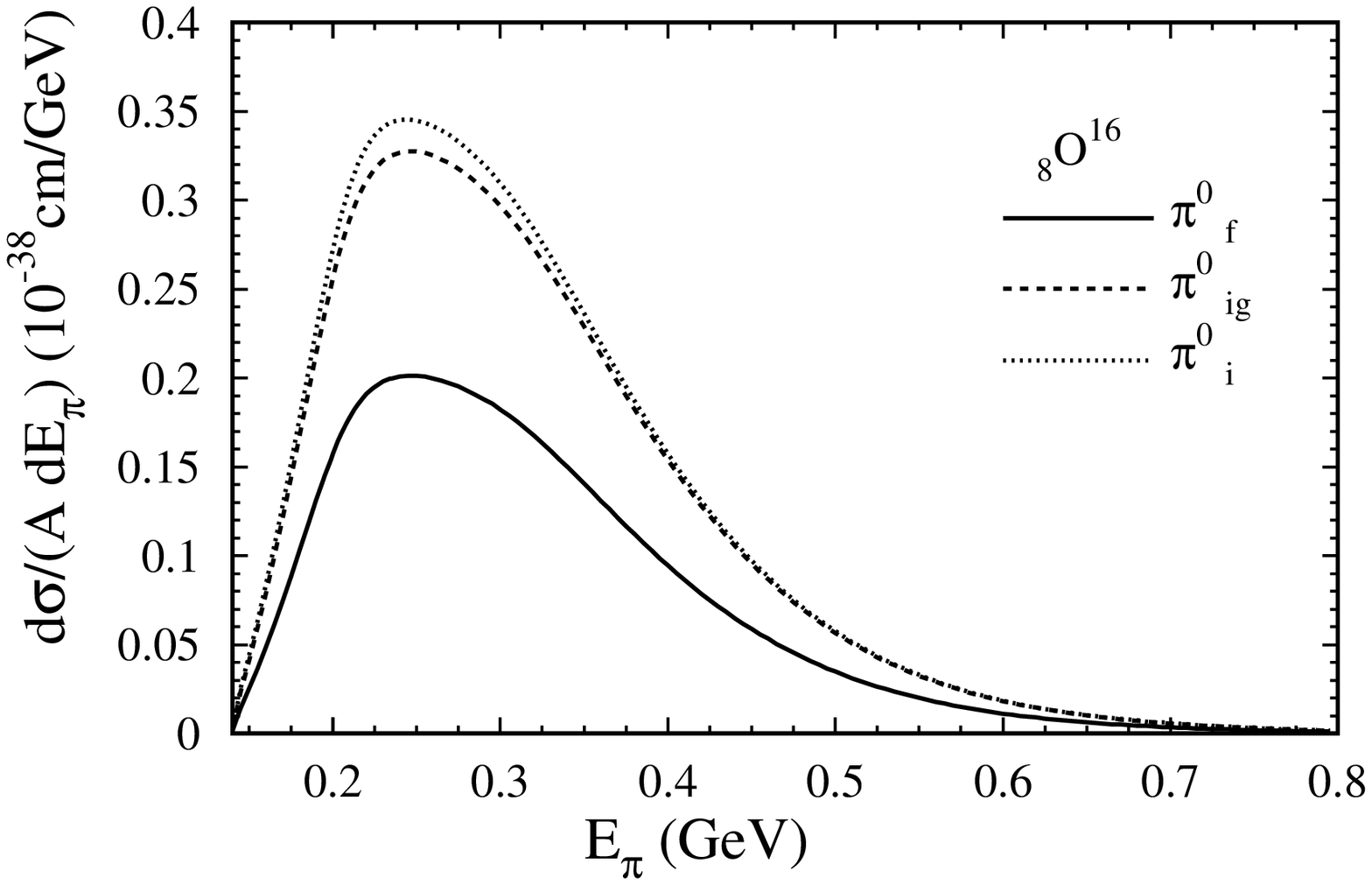}
\includegraphics[angle=0,width=8.9cm]{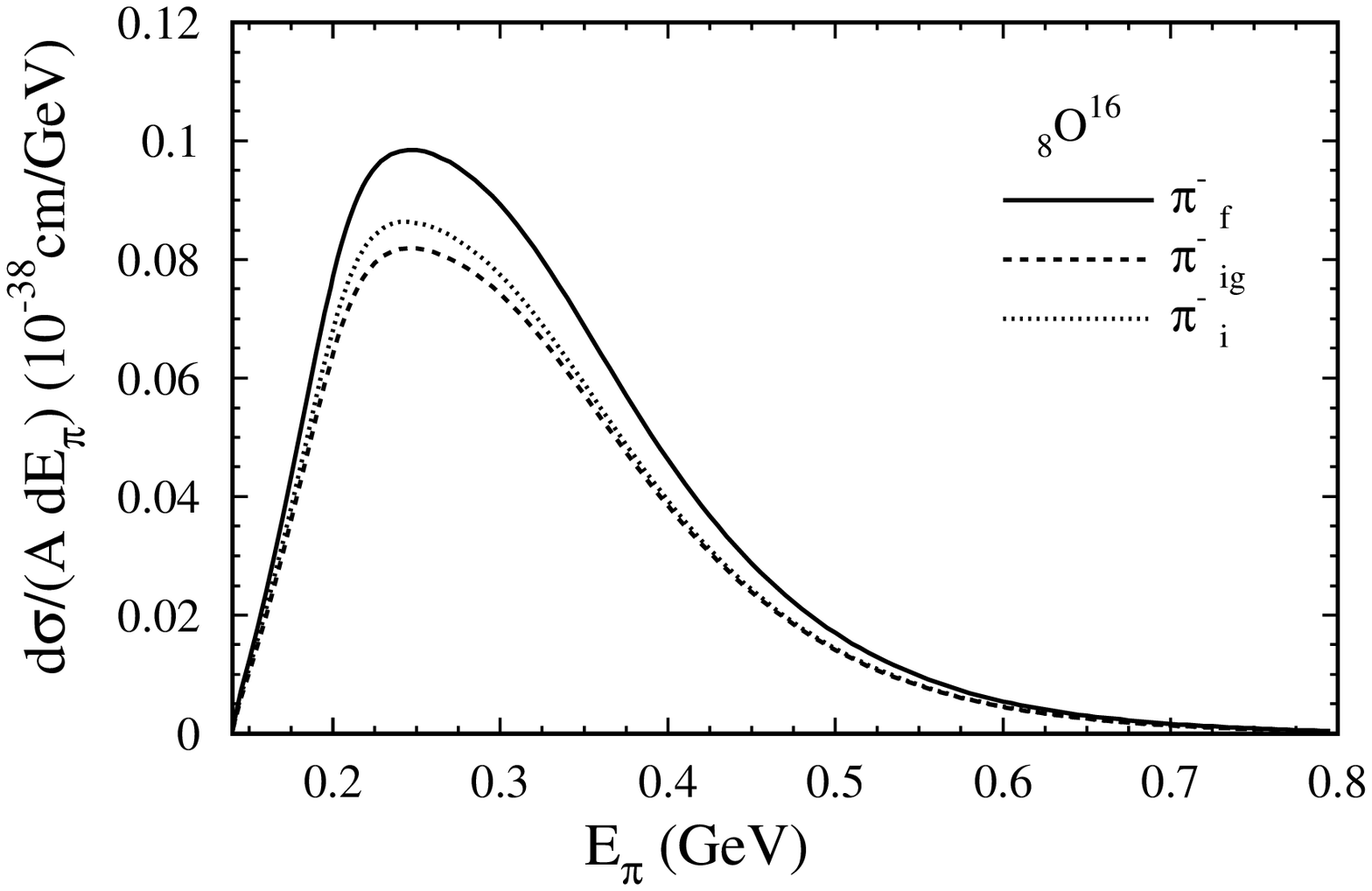}
\vspace*{-0.9cm}
\caption{\sf Differential cross section per nucleon for single pion spectra  
of $\pi^+,\ \pi^0, \ \pi^-$ for oxygen with $E_\nu = 1\ {\rm GeV}$
in dependence of pion energy $E_\pi$. The curves correspond to neutral current reactions.}
\label{fig:figoxygen}
\end{figure}

\begin{figure}[h]
\centering
\vspace*{-0.1cm}
\includegraphics[angle=0,width=8.9cm]{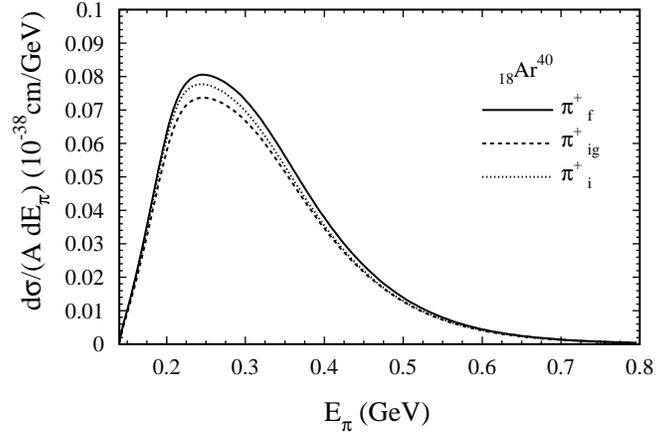}
\includegraphics[angle=0,width=8.9cm]{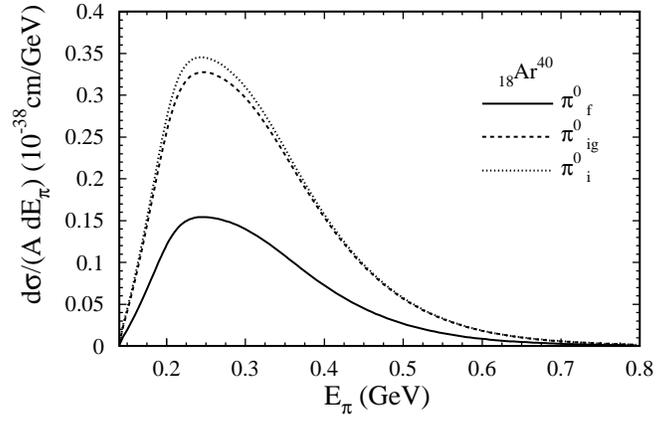}
\includegraphics[angle=0,width=8.9cm]{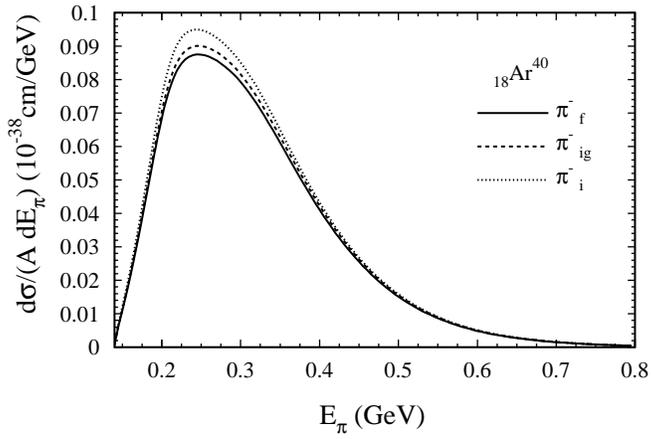}
\vspace*{-0.7cm}
\caption{\sf The same as in fig.\ \protect\ref{fig:figoxygen} for argon.}
\label{fig:figargon}
\end{figure}

\begin{figure}[h]
\centering
\vspace*{-0.1cm}
\includegraphics[angle=0,width=8.9cm]{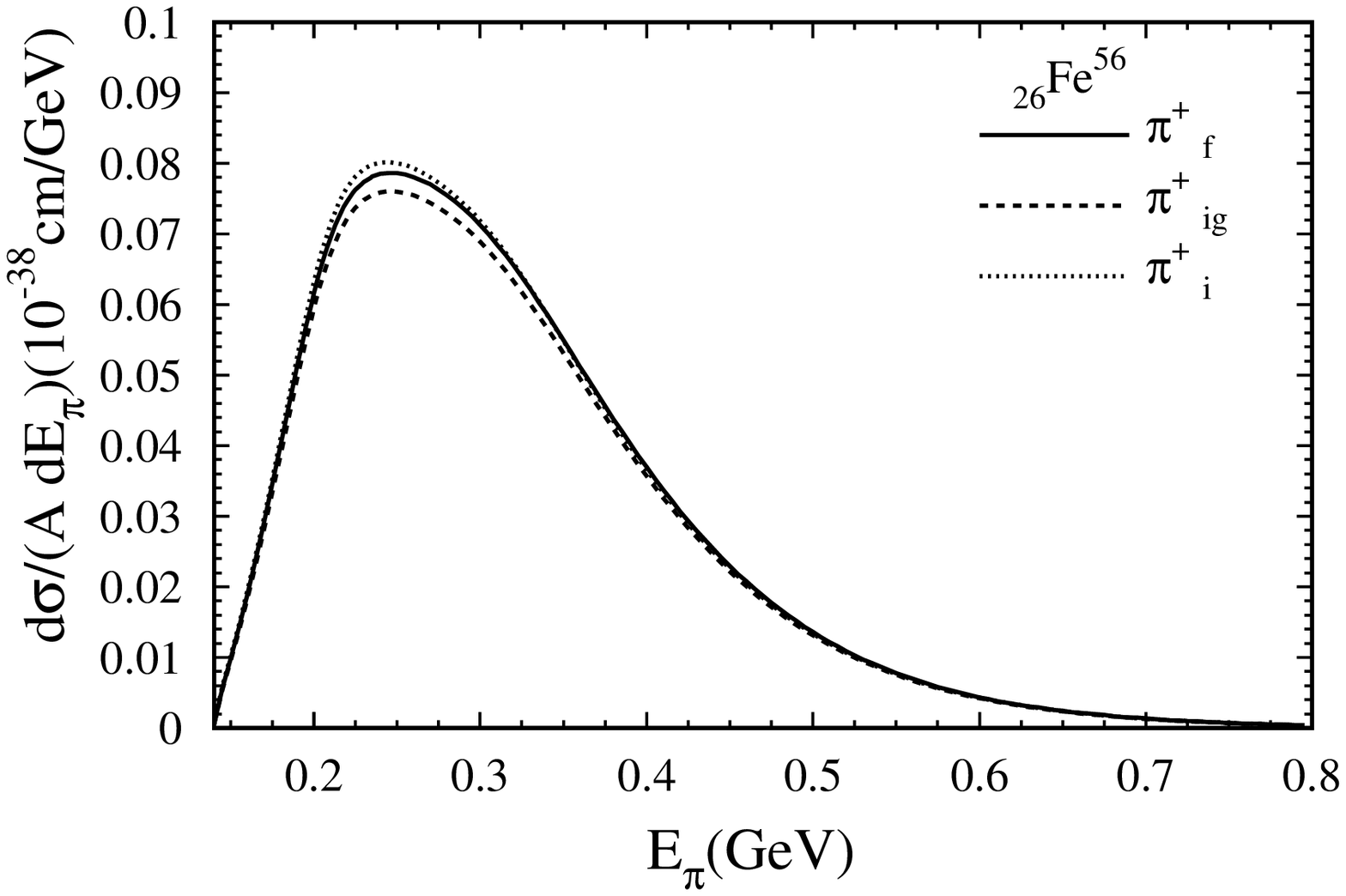}
\includegraphics[angle=0,width=8.9cm]{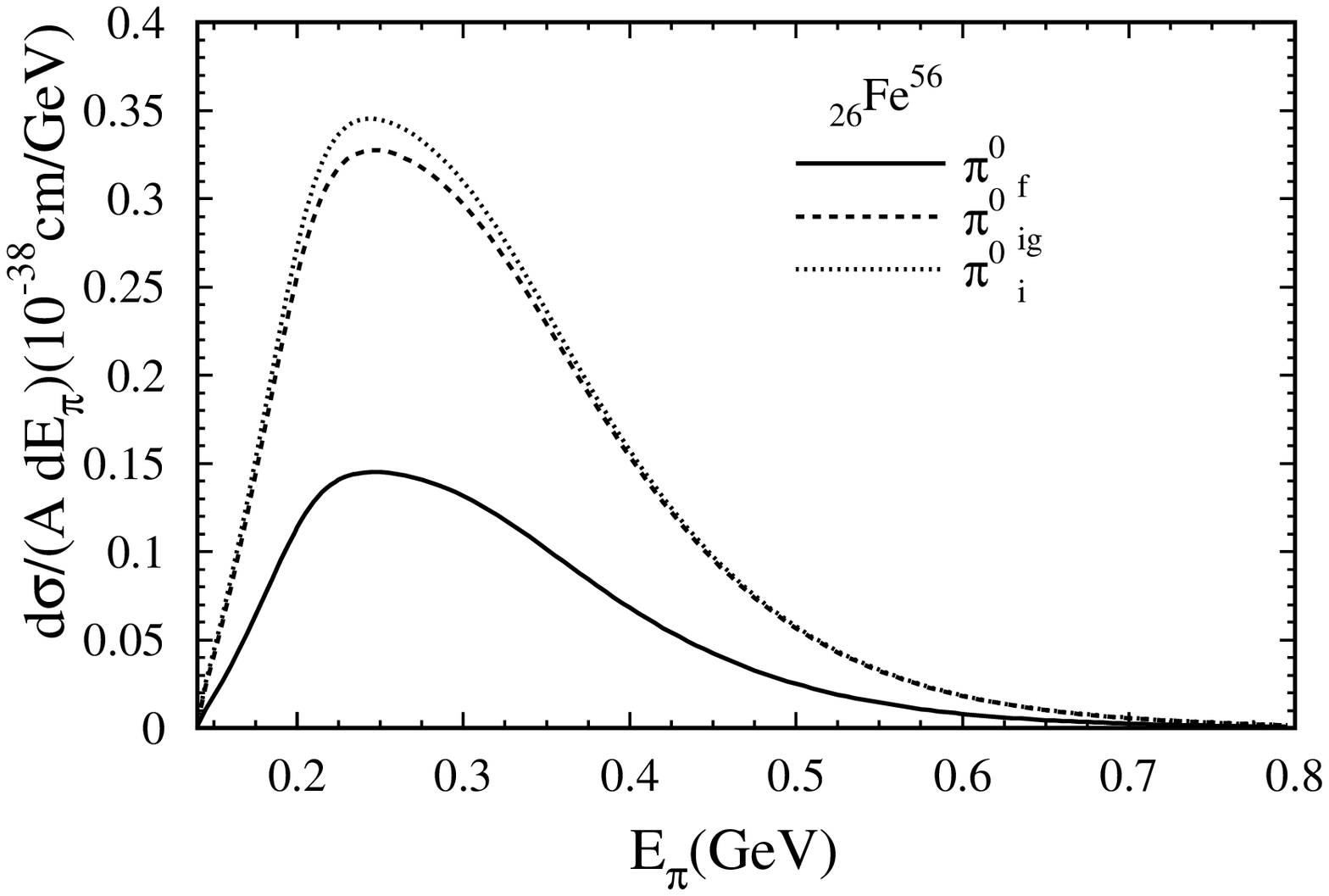}
\includegraphics[angle=0,width=8.9cm]{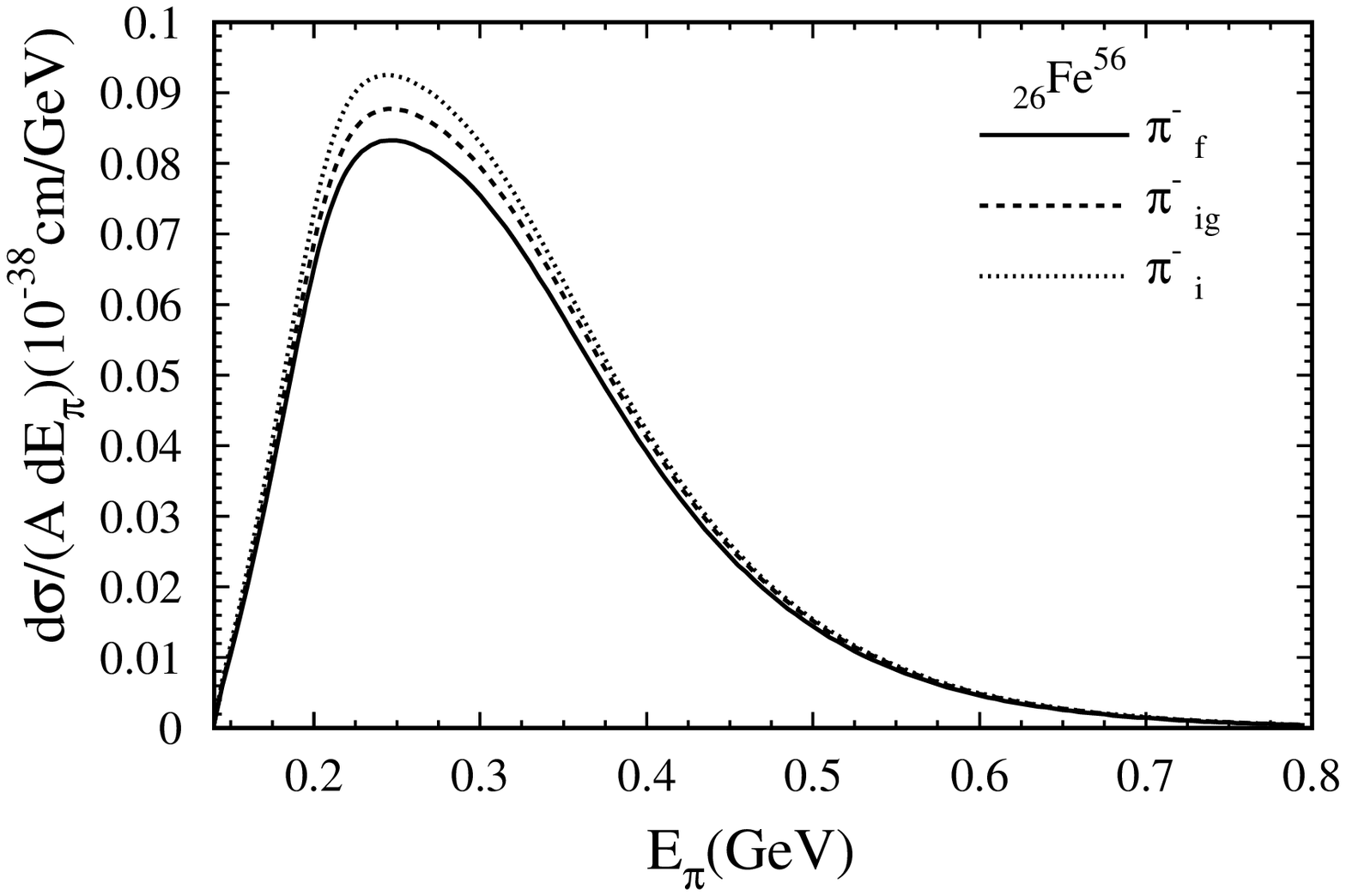}
\vspace*{-0.7cm}
\caption{\sf \sf The same as in fig.\ \protect\ref{fig:figoxygen} for iron.}
\label{fig:figiron}
\end{figure}

We begin with a discussion of the nuclear corrections
to the pion energy spectra in neutrino scattering shown
in Figs.\ \ref{fig:figoxygen}--\ref{fig:figiron}, 
where the curves are neutral current reactions.
The dotted lines are the spectra for the free nucleon 
cross sections.
The dashed lines include the effect of the Pauli suppression
(in step one of the two step process), whereas the solid line in addition
takes into account the pion multiple scattering.
These curves correct Figs.\ 8--16 in Ref.\ \cite{Paschos:2000be}.
Similar curves have been obtained recently by Leitner et al. \cite{Leitner:2006sp} 
who also noticed the error  in \cite{Paschos:2000be}.
Even though the models differ in the transport matrix, 
they both include charge exchange effects. For example, they both 
find that for reactions where the charge of the pions is the same with the charge 
of the current the pion yield shows a substantial decrease.
%
%where a phase space constraint had been overseen.
%[{\bf Comparison with Leitner et al.?; Mention that results for
%$W$ spectra remain unchanged?}]

\subsubsection{Electroproduction}
We now turn to the electroproduction.
To be specific, our analysis will be done under the
conditions of the 
Cebaf Large Acceptance Spectrometer (CLAS) 
at Jefferson Lab (JLAB).
The CLAS detector \cite{Mecking:2003zu}
covers a large fraction of the full solid angle with efficient
neutral and charged particle detection.
Therefore it is very well suited
to perform a high statistics measurement on various light and heavy 
nuclear targets and to test the ideas of pion multiple scattering
models.
% large acceptance -> dsigma/dQ^2dW ; phenomenological reconstruction
%
In the future these measurements can be compared with results
in neutrinoproduction from the Minerva experiment \cite{Drakoulakos:2004gn}
using the high intensity Numi neutrino beam.
If not stated otherwise we use an electron energy
$E_e = 2.7\ \gev$ in order to come as close as possible
to the relevant low energy range of the LBL experiments.
For the momentum transfer we take the values
$Q^2 = 0.4, 0.8\ \gevsq$
in order to avoid the experimentally and theoretically 
more problematic region at very low $Q^2$.
Results for larger $Q^2$ 
and larger energies, say $E_e = 10\ \gev$, are qualitatively 
very similar.

%\subsubsection{Comparison of exact results with the averaging approximation AV2}
%\FIGURE{
%%\centering
%\includegraphics[angle=0,width=6.5cm]{dq2dw_o_tot_0.3_pi0.eps}
%\includegraphics[angle=0,width=6.5cm]{dq2dw_o_tot_0.5_pi0.eps}
%\includegraphics[angle=0,width=6.5cm]{dq2dw_o_tot_0.3_pip.eps}
%\includegraphics[angle=0,width=6.5cm]{dq2dw_o_tot_0.5_pip.eps}
%%\vspace*{-2.0cm}
%\caption{\sf The double diffentical cross sections of one pion 
%electroproduction for an oxygen target  W is plotted for 
%$Q^2 = 0.3 \ {\rm GeV}$ and $Q^2 = 0.5 \ {\rm GeV}$ 
%fixing an electron energy $E_e = 2.7$ GeV. 
%The solid and dotted lines denote the cross section for 
%the exact ANP matrix $M(W;Q^2)$
%and the averaged approximation ANP matrix $\overline{\overline M}$
%respectively. The dashed line is the free .}
%\label{fig:2}
%}
\begin{figure}[h]
\centering
%\vspace*{-2.5cm}
\includegraphics[angle=0,width=7.0cm]{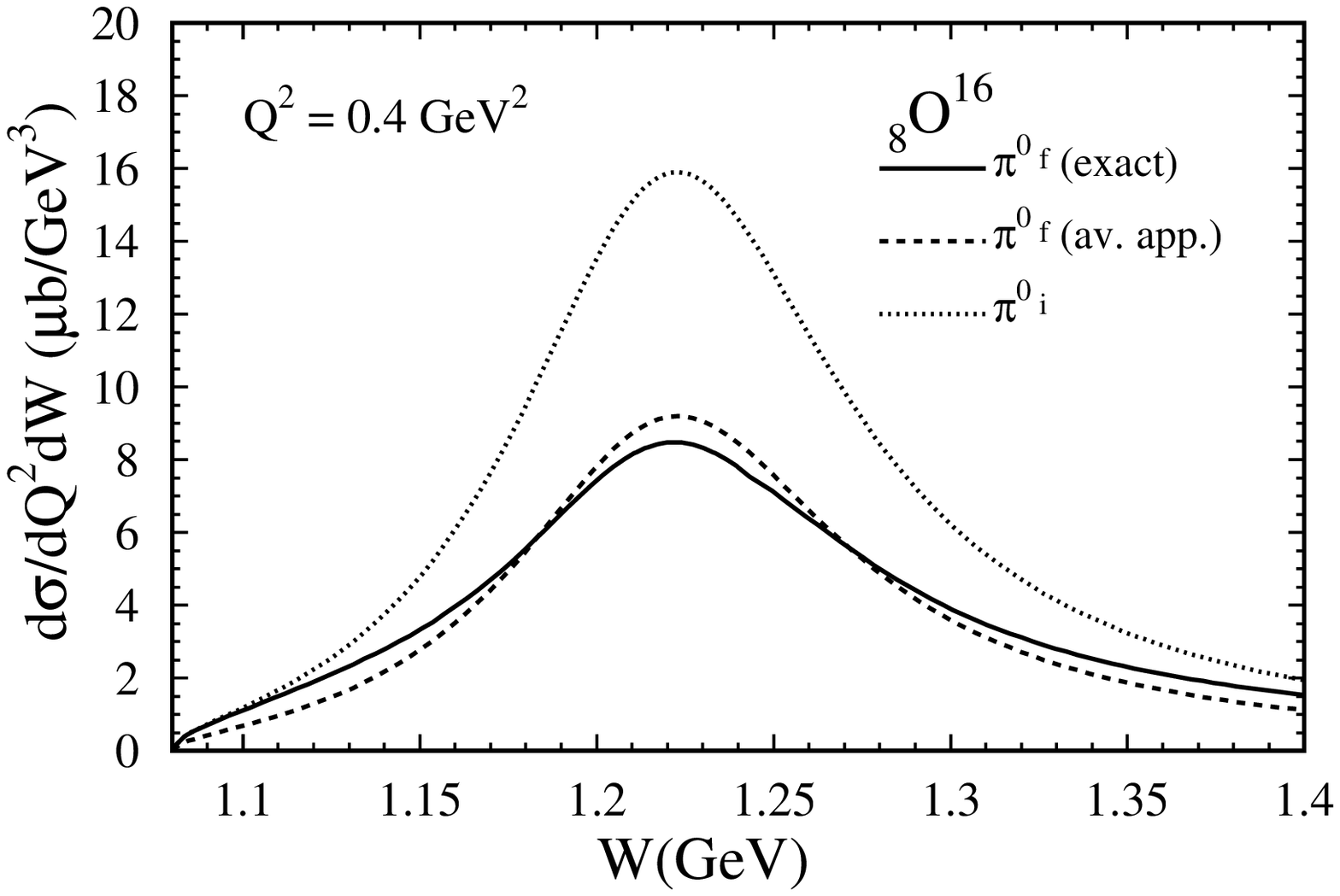}
\includegraphics[angle=0,width=7.0cm]{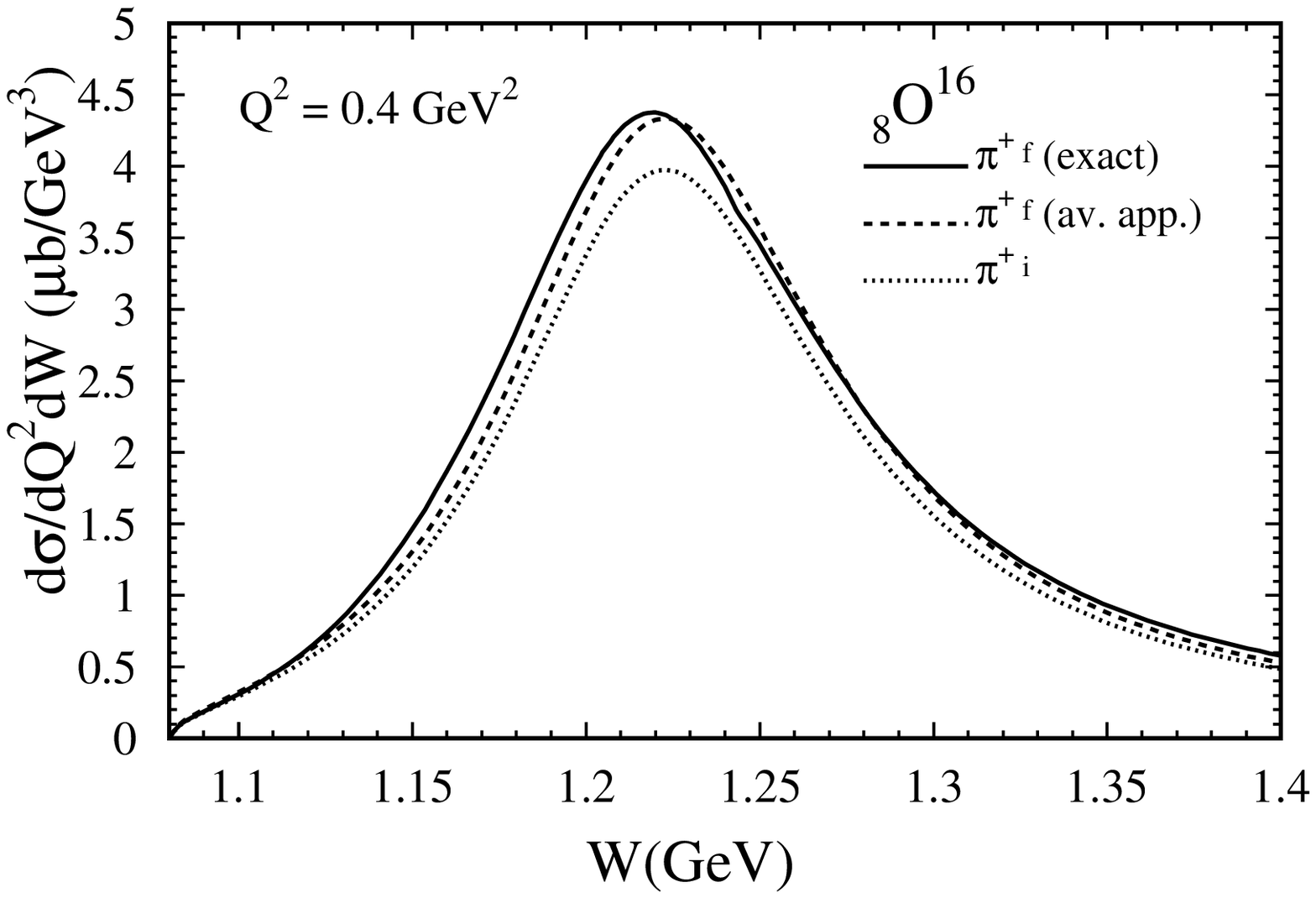}
\vspace*{-1.0cm}

\includegraphics[angle=0,width=7.0cm]{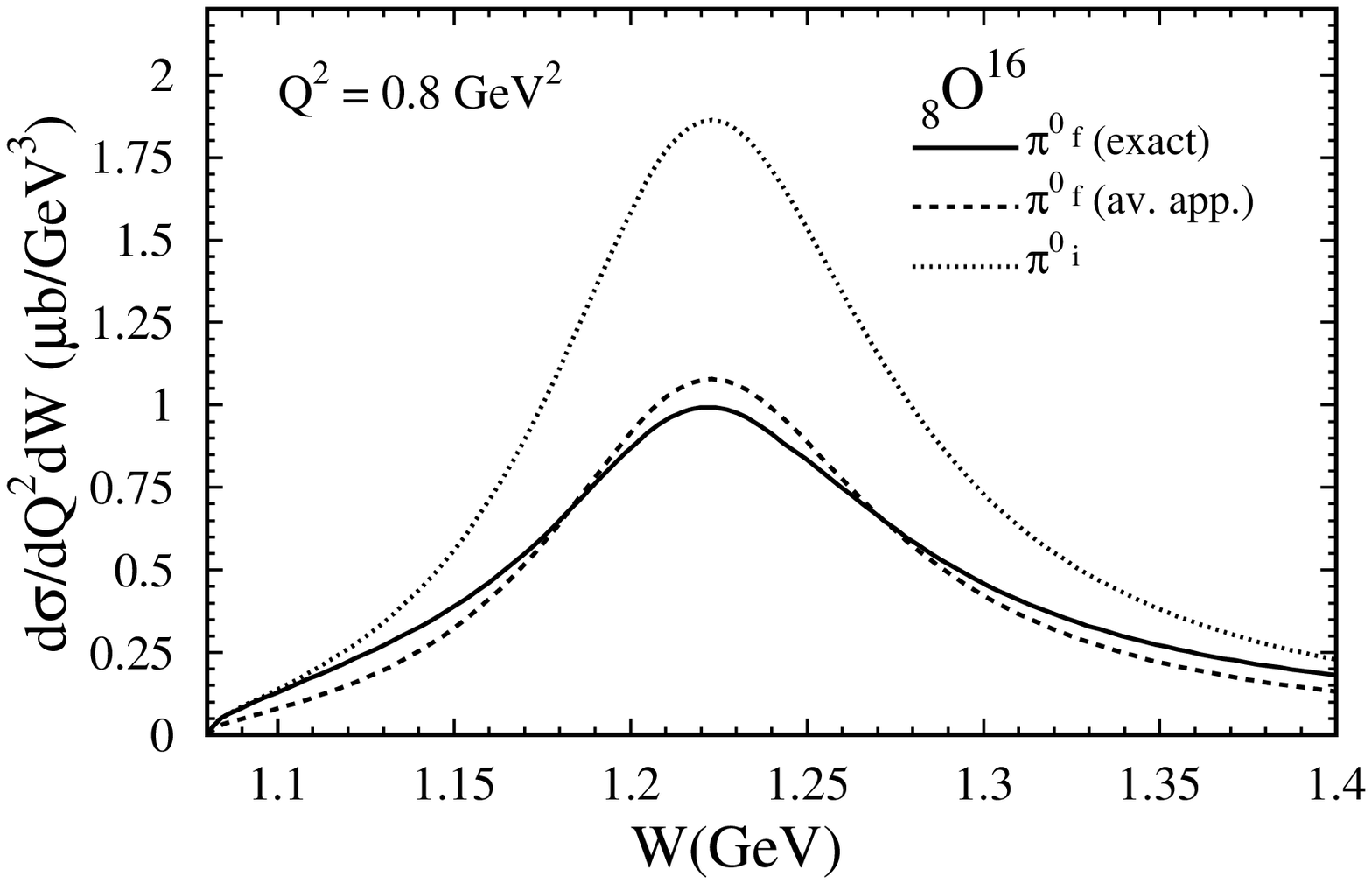}
\includegraphics[angle=0,width=7.0cm]{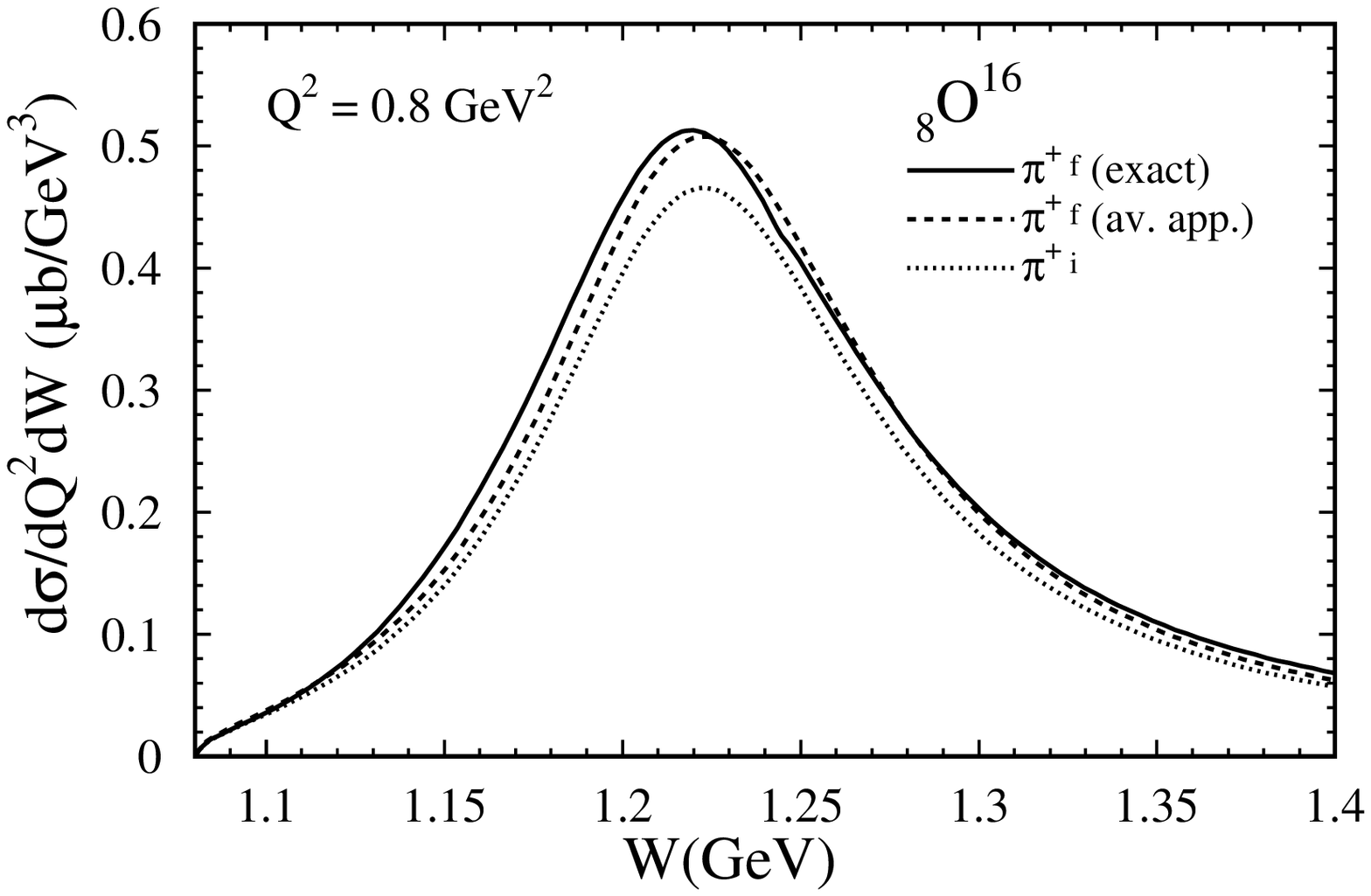}
\vspace*{-0.5cm}
\caption{\sf Double differential cross sections for single-pion 
electroproduction for an oxygen target in dependence of $W$.
Spectra for $\pi^0$ and $\pi^+$ production are shown 
for $Q^2 = 0.4\ \gevsq$ and $Q^2 = 0.8\ \gevsq$
using an electron energy $E_e = 2.7\ \gev$. 
The solid and dotted lines have been obtained 
according to \protect\eqref{eq:fac}
using the exact ANP matrix $M(W,Q^2)$
and \protect\eqref{eq:av2}
utilizing the double-averaged ANP matrix $\Mavav$ 
in \protect\eqref{eq:M1}, respectively. 
The dashed lines show the free nucleon 
cross section \protect\eqref{eq:free}.}
\label{fig:2}
\end{figure}

Figure \ref{fig:2} shows the double differential cross section 
$\der \sigma/\der Q^2 \der W$ for $\pi^+$ and $\pi^0$ production versus
%in dependence of 
$W$ for an oxygen target.
%The electron energy has been fixed to $E_e = 2.7\ \gev$ and we show
%results for momentum transfers 
%$Q^2= 0.4\ \gevsq$ and $Q^2=0.8\ \gevsq$, respectively.
%
The solid lines have been obtained with help of Eq.\ \eqref{eq:fac}
including the nuclear corrections.
The dashed lines show the result of the double-averaging approximation 
according to Eq.\ \eqref{eq:av2} using the ANP matrix in
Eq.\ \eqref{eq:M1}.
The dotted line is the free cross section in Eq.\ \eqref{eq:free}.
%
%As one can see, 
One sees,
the double-averaging approximation and 
the exact calculation give very similar results such that 
%the double averaging approximation 
the former
is well-suited for simple estimates 
to an accuracy of $10\%$ 
of pion rescattering effects.
We observe that the cross sections for $\pi^0$ production
are largely reduced by about $40\%$ due to the nuclear corrections.
This can be understood since the larger $\pi^0$ cross sections 
are reduced by absorption effects and charge exchange effects.
On the other hand, the $\pi^+$ cross sections are even slightly enlarged,
because the reduction due to pion absorption is compensated by
an increase due to charge exchange.
The compensation is substantial since the $\pi^0$ yields are dominant.
%\TODO{Include in figure comparison: exact with and without Fermi motion ???}

\begin{figure}[h]
\centering
%\vspace*{-2.5cm}
\includegraphics[angle=0,width=7.0cm]{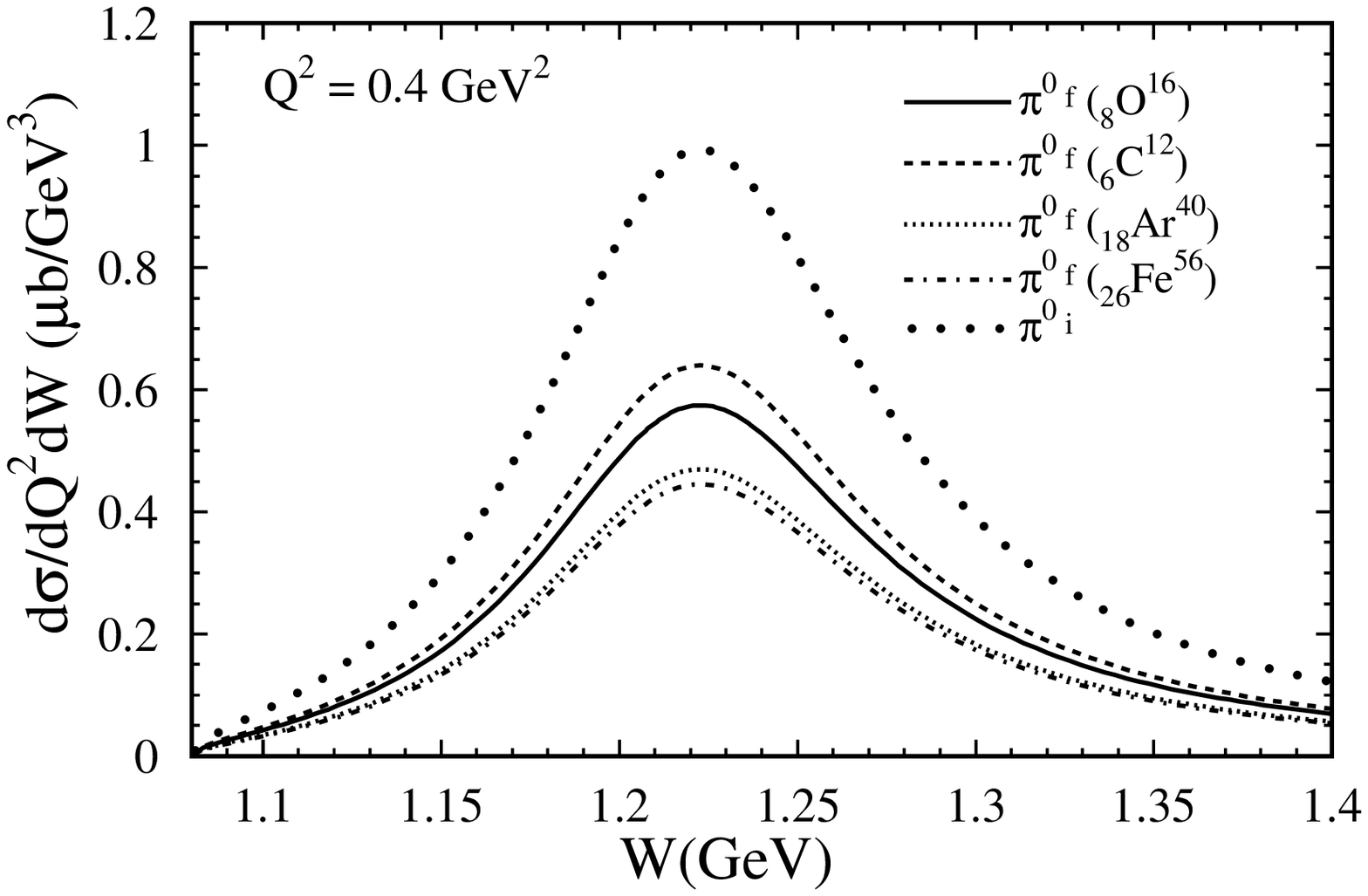}
\includegraphics[angle=0,width=7.0cm]{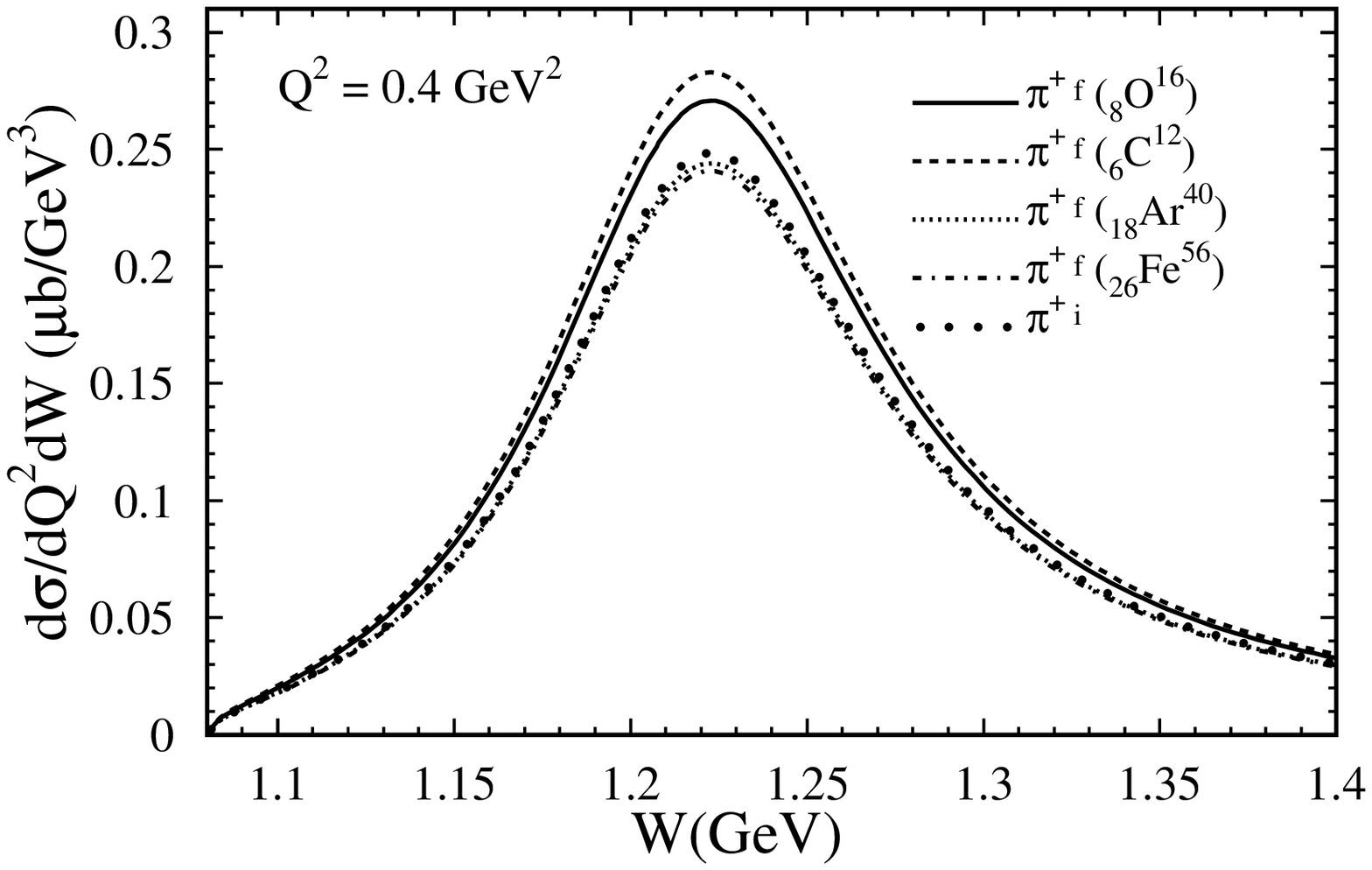}
%\vspace*{-1.0cm}

%\includegraphics[angle=0,width=7.0cm]{dq2dw_o_tot_0.8_pi0.eps}
%\includegraphics[angle=0,width=7.0cm]{dq2dw_o_tot_0.8_pip.eps}
\vspace*{-0.5cm}
\caption{\sf Double differential cross sections per nucleon 
for single-pion electroproduction for 
different target materials.
$W$-spectra for $\pi^0$ and $\pi^+$ production are shown 
for $Q^2 = 0.4\ \gevsq$ using an electron energy $E_e = 2.7\ \gev$. 
The pion rescattering corrections have been calculated in 
the double-averaging approximation \protect\eqref{eq:av2} using
the ANP matrices in \protect\eqref{eq:M1} and App.\ \protect\ref{app:M1}.
For comparison, the free nucleon cross section \protect\eqref{eq:free} 
is shown.}
\label{fig:targets}
\end{figure}

%\subsubsection{Results for various target materials}
In Fig.\ \ref{fig:targets} 
double differential cross sections per nucleon 
for different target materials are presented.
The electron energy and the momentum transfer have been chosen as
$E_e = 2.7\ \gev$ and $Q^2= 0.4\ \gevsq$, respectively.
The results for the pion rescattering corrections have been obtained
within the double-averaging approximation \eqref{eq:av2} 
which allows for a simple
comparison of the dependence on the target material in terms of the
matrices $\Mavav[T]$ which can be found in
Eq.\ \eqref{eq:M1} and App.\ \ref{app:M1}. 
For comparison the free nucleon cross section \eqref{eq:free}
(isoscalar $\tfrac{p+n}{2}$) is also shown.
As expected, the nuclear corrections become larger with increasing
atomic number from carbon to iron.

%\subsubsection{Dependence on pion absorption}
One of the input quantities for calculating the transport function
$f(\lambda)$ in the ANP model is the pion absorption cross section 
$\sigabs(W)$
describing the probability that the pion is absorbed in a 
single rescattering process.
For $\sigabs(W)$ the ANP article reported results for
%model employs 
two parameterizations,
models A and B,
taken from Refs.\ \cite{Sternheim:1972ad,Silbar:1973em} which have very
%exhibit quite a
different $W$-dependence and normalization.
However, the predictions of the ANP model in the 
double-averaging approximation are primarily
%mainly 
sensitive to the 
normalization of the pion absorption cross section at 
$W \simeq m_\Delta$ \cite{Schienbein:2003xy}. 
Using data by Merenyi et al.\ \cite{Merenyi:1992gf} for a neon target
it was found that about $25\% \pm 5\%$ of pions are absorbed 
%in neon \cite{Schienbein:2003xy} 
%allowing to fix 
making possible the determination of 
the normalization of
$\sigabs(W)$
%the pion absorption cross section 
with a $20\%$ accuracy.

\begin{figure}[h]
\centering
%\vspace*{-2.5cm}
\includegraphics[angle=0,width=7.0cm]{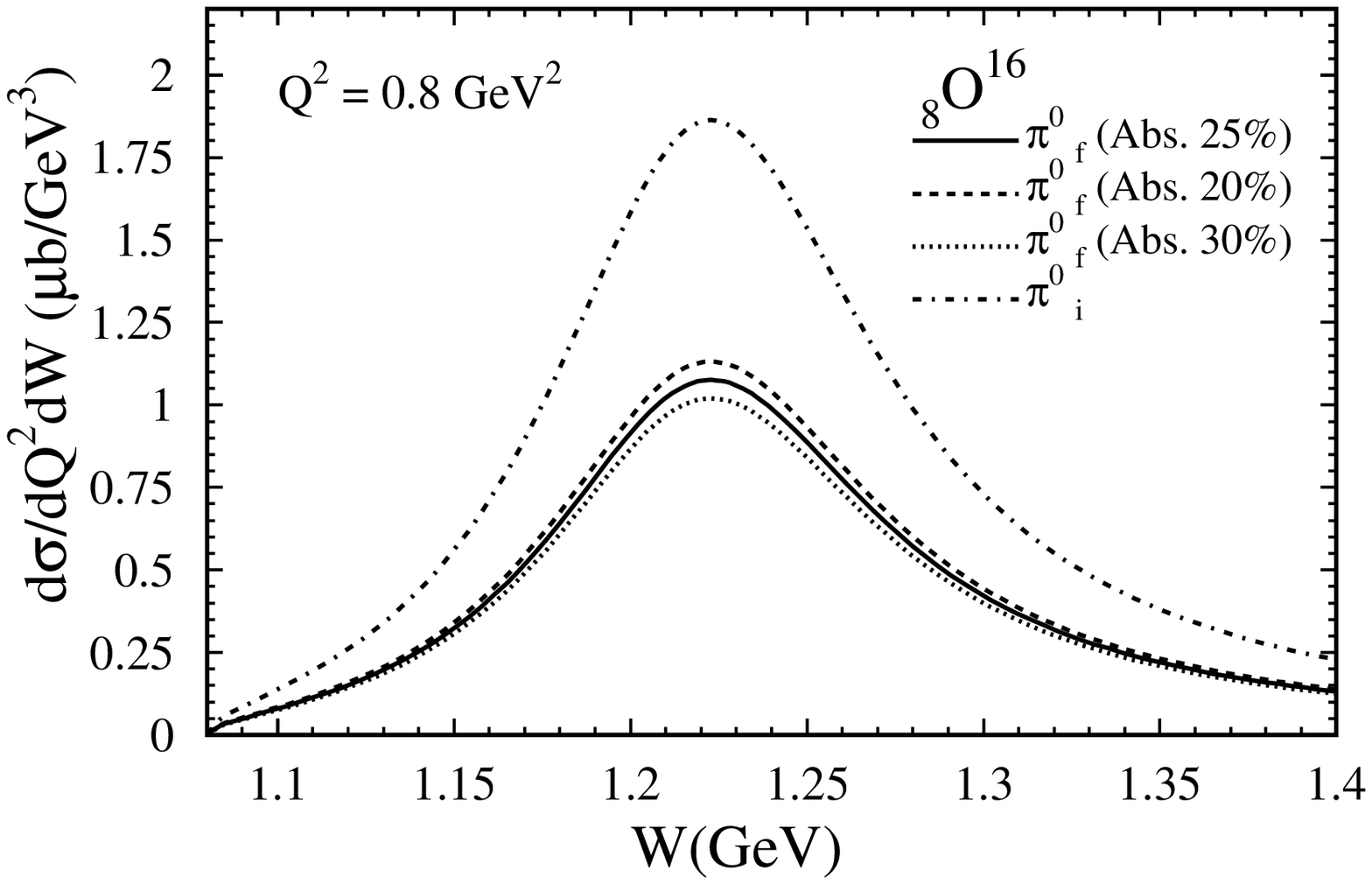}
\includegraphics[angle=0,width=7.0cm]{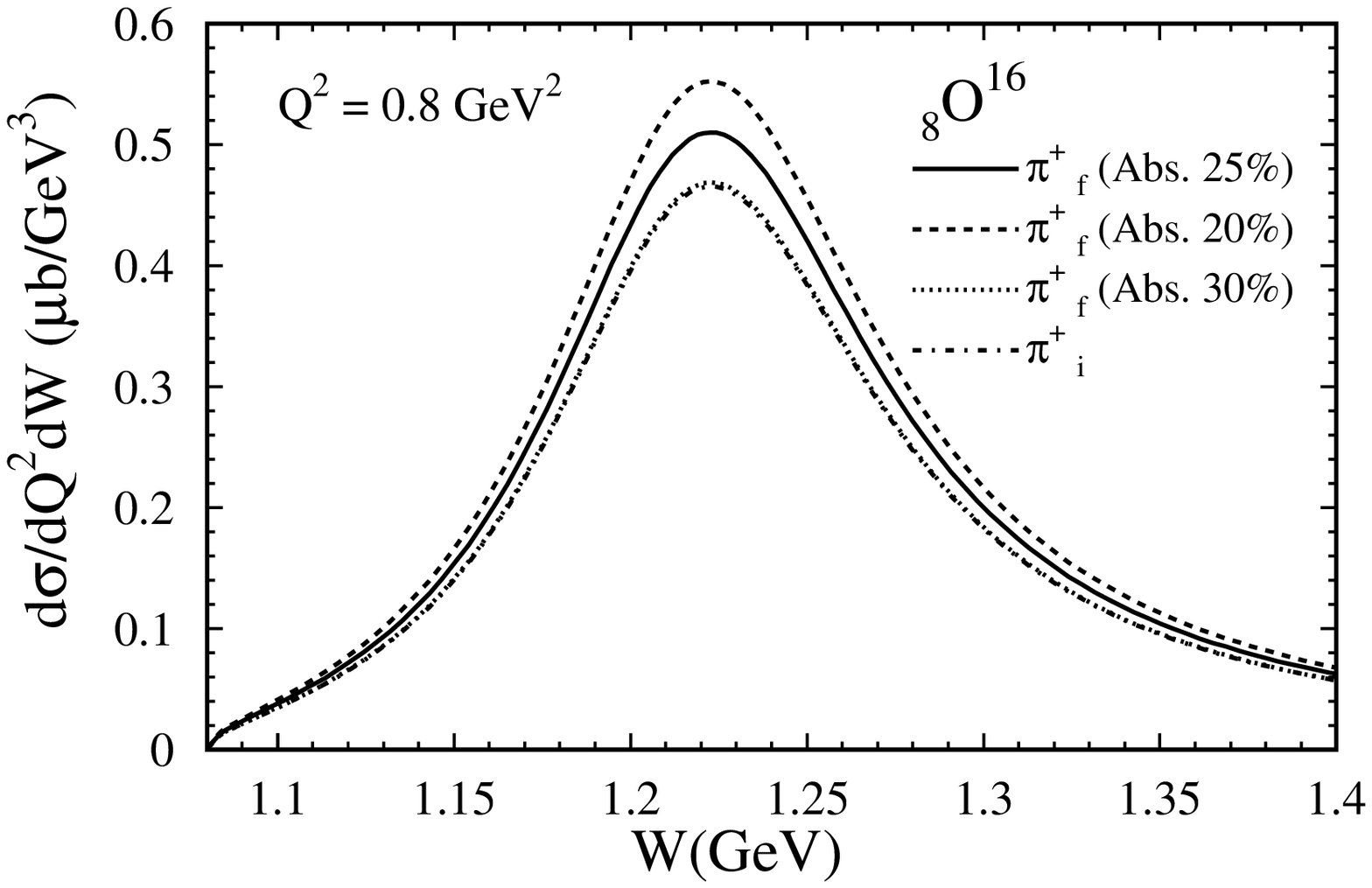}
%\vspace*{-1.0cm}

%\includegraphics[angle=0,width=7.0cm]{dq2dw_o_tot_0.5_pi0.eps}
%\includegraphics[angle=0,width=7.0cm]{dq2dw_o_tot_0.5_pip.eps}
\vspace*{-0.5cm}
\caption{\sf Double differential cross sections per nucleon 
for single-pion electroproduction for oxygen with $20\%$ (dashed line),
$25\%$ (solid line) and $30\%$ (dotted line) pion absorption.
Furthermore, $Q^2 = 0.8\ \gevsq$ and $E_e = 2.7\ \gev$.
The $\pi^0$ and $\pi^+$ spectra have been calculated in
the double-averaging approximation \protect\eqref{eq:av2}
utilizing the matrices in App.\ \protect\ref{app:M2}. 
For comparison, the free nucleon cross section \protect\eqref{eq:free} 
is shown as well.}
\label{fig:abs1}
\end{figure}

In order to investigate the theoretical uncertainty due
to pion absorption effects we show in 
Fig.\ \ref{fig:abs1} double differential cross sections 
$\der \sigma/\der Q^2 \der W$ for $\pi^+$ and $\pi^0$ production vs
%in dependence of 
$W$ for different amounts of pion absorption in oxygen:
$25\%$ (solid line), $20\%$ (dashed line), $30\%$ (dotted line).
The $\pi^0$ and $\pi^+$ spectra have been calculated in
the double-averaging approximation \eqref{eq:av2}
utilizing the matrices in App.\ \ref{app:M2}. 
The three curves represent the theoretical uncertainty due to pion 
absorption effects.
For comparison, the free nucleon cross section \eqref{eq:free} 
is shown as well.
%\TODO{The same for carbon and iron ???}

Although the predictions of the ANP model are mainly sensitive
to $\sigabs(W \simeq m_\Delta)$ it would be interesting
to obtain more information on the detailed $W$-shape.
The fraction of absorbed pions can be determined 
by measuring the inclusive pion production cross sections 
for a nuclear target divided by the free nucleon cross sections,
\begin{equation}
\ABS(Q^2,W) =     1 - 
  \frac{\sum_{k=0,\pm} \frac{\der \sigma(_ZT^A;\pi^k)}{\der Q^2 \der W}}
         {\sum_{j=0,\pm} \frac{\der \sigma(N_T;\pi^j)}{\der Q^2 \der W}}
= 1 - A_p(Q^2,W)
\, ,
\label{abs}
\end{equation}
where $A_p$ has been introduced in \eqref{eq:M}.
This quantity is related to $\sigabs(W)$ as can be
seen by linearizing the transport function $f(\lambda,W)$
\cite{Paschos:2004qh,Schienbein:2003xy}
\begin{equation}
\ABS(Q^2,W) \simeq \frac{1}{2} \bar{L} \rho_0 \times \sigabs(W)\ .
\label{eq:sigabs}
\end{equation}
Here $\bar{L}$ is the effective length of the nucleus 
averaged over impact parameters and $\rho_0$ the charge density in the
center. As an example, for oxygen one finds $\bar{L} \simeq 1.9 R$ 
with radius $R \simeq 1.833\ {\rm fm}$ and 
$\rho_0 = 0.141\ {\rm fm}^{-3}$.
Therefore, the $W$-dependence of $\sigabs(W)$ can be reconstructed
from the fraction of absorbed pions, i.e.\ $\ABS(Q^2,W)$.
Summing over the three charged pions eliminates charge exchange effects.
%
%Therefore, the ANP model predicts 
%the $W$-dependence of the fraction of 
%absorbed pions, $\ABS(Q^2,W)$, to vary only mildly between 
%different target materials used in LBL experiments.
%This expectation can be verified by measuring $\ABS(Q^2,W)$ for 
%target materials like carbon, oxygen, argon and iron.

\begin{figure}[htb]
\centering
\vspace*{-1cm}
\includegraphics[angle=0,width=7.9cm]{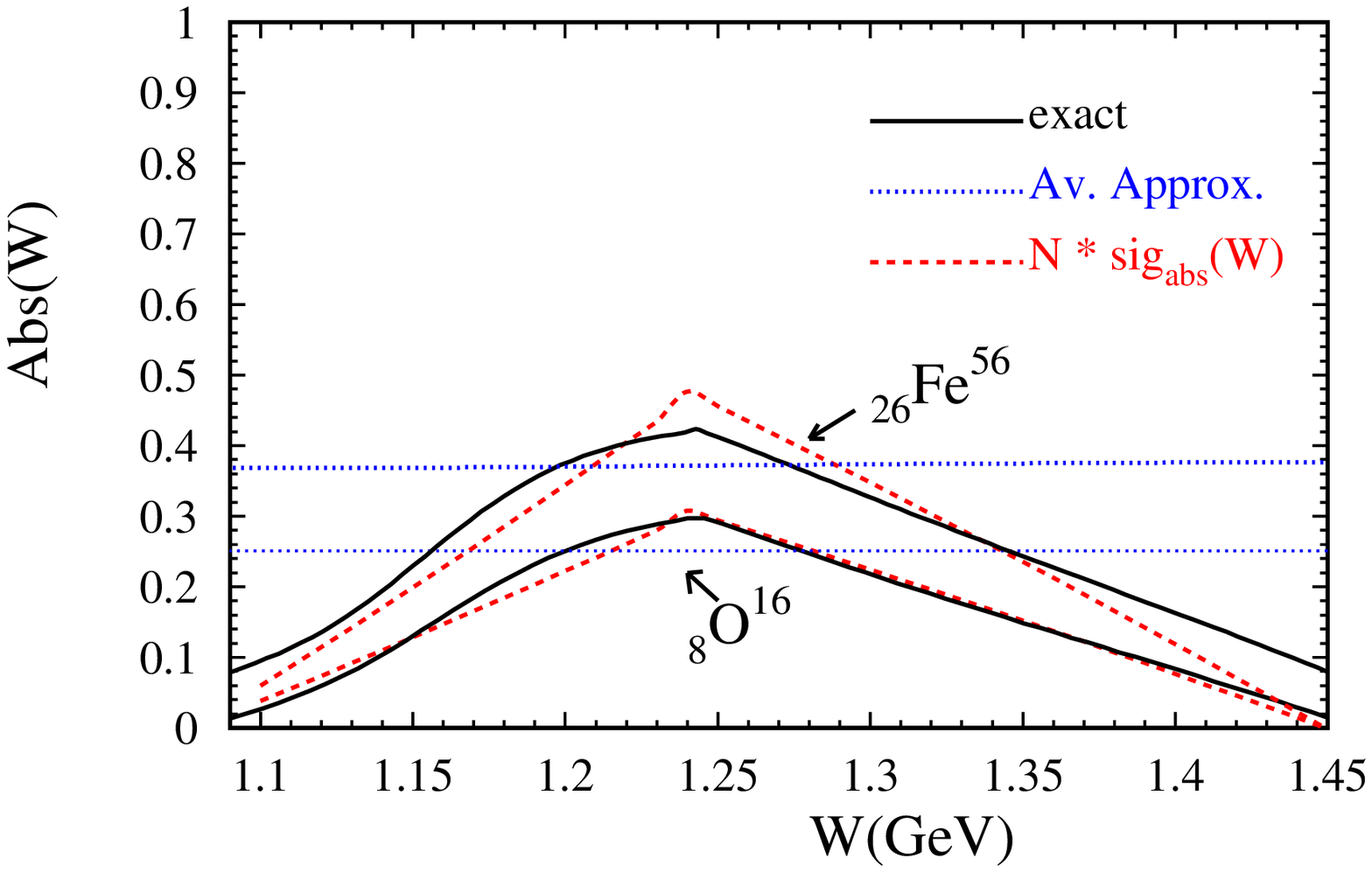}
\vspace*{-0.8cm}
\caption{\sf The fraction of absorbed pions, $\ABS(Q^2,W)$, in
dependence of $W$ for oxygen and iron targets for $Q^2=0.3\ \gevsq$.
Also shown is the cross section $\sigabs(W)$ (model B) multiplied by
free normalization factors (dashed lines).
The dotted lines are the result for $\ABS(Q^2,W)$ 
in the averaging approximation.}
\label{fig:sigabs}
\end{figure}

In order to verify the linearized approximation
%simple relation 
in Eq.\ \eqref{eq:sigabs},
we show
in Fig.\ \ref{fig:sigabs} the ANP model prediction 
for $\ABS(Q^2,W)$ for oxygen and iron targets with $Q^2 = 0.3\ \gevsq$.
This prediction strongly depends on the shape of the cross section $\sigabs(W)$ for
which we use model B from Refs.\ \cite{Silbar:1973em}.
$\sigabs(W)$ multiplied by a free
normalization factors for oxygen and iron, respectively, 
is depicted by the dashed lines.
Obviously, Eq.\ \eqref{eq:sigabs} is quite 
well satisfied for oxygen and still reasonably good for iron.
Finally, the dotted line shows the result of the averaging approximation.
We conclude that $\sigabs(W)$ can be extracted with help of Eqs. \eqref{abs} and \eqref{eq:sigabs}.

For completeness, we mention that 
the pion absorption in nuclei is reported in various articles 
\cite{Ashery:1987nt,*Ingram:2001xs,*Ransome:2005vb}.
For comparisons one should be careful because the absorption cross
sections in pi-nucleus and in neutrino-nucleus reactions are different, 
in the former case it is a surface effect while in the latter it occurs
everywhere in the nucleus.

\begin{figure}[htb]
\centering
\vspace*{-1cm}
\includegraphics[angle=0,width=7.9cm]{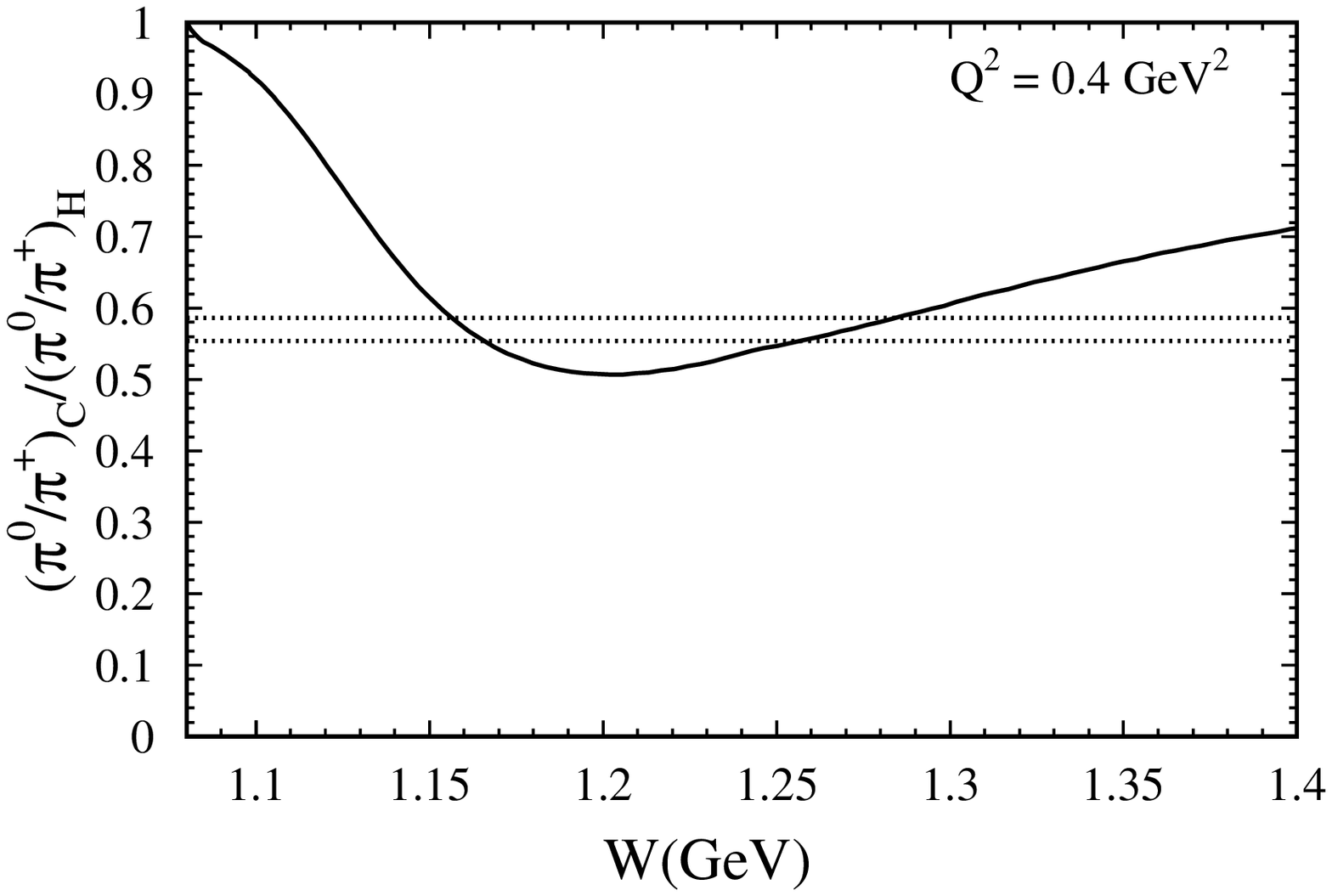}
\vspace*{-0.8cm}
\caption{\sf Double ratio of single pion electroproduction cross sections
in dependence of $W$ for fixed $Q^2=0.4\ \gevsq$ 
as defined in Eq.\ \protect\eqref{eq:doubleratio}. 
The dotted lines show results in the double averaging approximation
with varying amounts of absorption.}
\label{fig:doubleratio}
\end{figure}

%\subsubsection{Double ratio}
A useful test of charge exchange effects is provided by the
double ratio
\begin{equation}
{\rm DR}(Q^2,W) = \left(\frac{\pi^0}{\pi^+ + \pi^-}\right)_A / 
\left(\frac{\pi^0}{\pi^+ + \pi^-}\right)_p
\label{eq:doubleratio}
\end{equation}
where $(\pi^i)_A$ represents the doubly differential cross section $\der \sigma/\der Q^2 \der W$
for the production of a pion $\pi^i$ in $eA$ scattering.
This observable is expected to be rather robust with respect to radiative corrections
and acceptance differences between neutral and charged pions.\footnote{We are 
grateful to S.\ Manly for drawing our attention to the double ratio.}
In Fig.\ \ref{fig:doubleratio} we show the double ratio for a carbon target in
dependence of $W$ for a fixed $Q^2 = 0.4\ \gevsq$.
The dependence on $Q^2$ is weak and results for other values of $Q^2$ are very similar.
The solid line shows the exact result, whereas the dotted lines have been obtained
in the double averaging approximation with minimal and maximal amounts of
pion absorption. As can be seen, the results are rather insensitive to the
exact amount of pion absorption.
Without charge exchange effects (and assuming similar absorption of charged and neutral pions)
the double ratio would be close to unity.
As can be seen, the ANP model predicts a double ratio smaller than $0.6$ in the
region $W \simeq 1.2\ \gev$.
A confirmation of this expectation would be a clear signal of pion charge exchange 
predominantly governed by isospin symmetry.
In this case it would be interesting to go a step further and to study similar
ratios for pion angular distributions.

\section{Summary}
\label{sec:summary}
Lepton induced reactions on medium and heavy nuclei include the 
rescattering of produced pions inside the nuclei.
This is especially noticeable in the $\Delta$-resonance region, where 
the produced resonance decays into a nucleon and a pion.
In the introduction and section \ref{sec:sec2} we reviewed the progress that 
has been made in the calculations of neutrino-induced reactions 
on free protons and neutrons, because we needed them for following calculations.
For several resonances the vector form factors have been recently determined by using 
electroproduction results in Jefferson Laboratory \cite{Lalakulich:2006sw}.
For the axial form factors modified dipoles give an accurate description of the data.
For the purposes of this article (studies of nuclear corrections) 
it suffices to deduce the  electroproduction cross sections through Eqs. \eqref{eq:em}
and \eqref{eq:em1}.

The main contribution of this article is contained in section \ref{sec:sec3},
where we describe important features of the ANP model and define single- 
and double averaged transport matrices.
Two important aspects of rescattering are emphasized:
(i) the absorption of the pions and (ii) charge exchange occurring 
in the multiple scattering, where we have shown that special features of the 
data are attributed to each of them. Finally we propose specific ratios of electroproduction
reactions that are sensitive to the absorption cross section and to charge exchange effects.

Using the model we calculate the transport matrix for various absorption
cross sections and nuclei and present the results in appendix \ref{app:M1}. 
We also calculated the pion energy spectra with and without nuclear corrections.
The results appear in figures \ref{fig:figoxygen}--\ref{fig:figiron} and can be compared with other 
calculations \cite{Leitner:2006sp}. Comparison of the double averaged approximation 
with the exact ANP calculation shows small differences (figure \ref{fig:2}).
As mentioned already, 
electroproduction data are very useful in testing several aspects of the model and its predictions.
For the absorption cross section we propose in  Eq.\ \eqref{abs} a ratio 
that depends only on ${\rm A_p}(Q^2,W)=g(Q^2,W) f(1,W)$.
Since we consider isoscalar targets and sum over the charges of the pions, 
charge exchange terms are eliminated.
This leaves over the dependence on charge independent effects, 
like the Pauli factor and the average absorption;
this is indeed the average absorption of pions and even includes the absorption 
of the $\Delta$-resonance itself.

Another ratio $({\rm DR}(Q^2,W))$ is sensitive to charge exchange effects.
In the double ratio the dependence on ${\rm A_p}(Q^2,W)$ drops out 
and the surviving terms are isospin dependent.
Our calculation shows that the ratio depends on $W$ with the largest reduction
occurring in the region $1.1 < W <1.25$ GeV.
Finally, the $\Delta(1232)$ is a sharply peaked resonance,
where the resonant interaction, takes place over small ranges of the kinematic variables,
so that averaging over them gives accurate approximations.
This is analogous to a narrow width approximation. 
%In several comparison 
%in this article confirm the expectation that average quantities give rather 
%accurate approximations of more extensive calculations.
Several comparisons 
in this article confirm the expectation that averaged quantities give rather 
accurate approximations of more extensive calculations.

%\\\\
\vspace*{2cm}
\noindent{\bf Acknowledgments}

\noindent
We wish to thank W.\ Brooks and S.\ Manly for many useful discussions,
their interest and encouragement.
The work of J.\ Y.\ Yu is supported by 
the Deutsche Forschungsgemeinschaft (DFG) through Grant No.\ YU 118/1-1.

\newpage
\noindent{\bf Appendix}
\begin{appendix}
\section{Charge exchange matrices in the double averaging approximation}
\label{app:M1}
\underline{Carbon:}\\
\begin{equation}
\Mavav(_{6}C^{12}) = \overline{\overline{A_p}}\left(\begin{array}{ccc}
{0.826}& {0.136} & {0.038} \\
{0.136}& {0.728} & {0.136} \\
{0.038}& {0.136} & {0.826}
\end{array}\right)
\label{eq:MC}
\end{equation}
with $\overline{\overline{A_p}} =0.791$ .\\

\noindent\underline{Argon:}\\
\begin{equation}
\Mavav(_{18}Ar^{40}) = \overline{\overline{A_p}}\left(\begin{array}{ccc}
{0.733}& {0.187} & {0.080} \\
{0.187}& {0.626} & {0.187} \\
{0.080} & {0.187} & {0.733}
\end{array}\right)
\label{eq:MAr}
\end{equation}
with $\overline{\overline{A_p}} =0.657$ .

\noindent \underline{Iron:}\\
\begin{equation}
\Mavav(_{26}Fe^{56}) = \overline{\overline{A_p}}\left(\begin{array}{ccc}
{0.720}  &  {0.194}& {0.086} \\
{0.194} & {0.613} & {0.194} \\
{0.086} & {0.194} & {0.720}
\end{array}\right)
\label{eq:MFe}
\end{equation}
with $\overline{\overline{A_p}} =0.631$ .

\section{Charge exchange matrices for various amounts of pion absorption}
\label{app:M2}
\underline{Carbon:} \\

15\% absorption
\begin{equation}
\Mavav(_{6}O^{12}) = \overline{\overline{A_p}}\left(\begin{array}{ccc}
{0.817} & {0.141} & {0.041} \\
{0.141} & {0.718} & {0.141} \\
{0.041} & {0.141} & {0.817}
\end{array}\right)
\label{eq:MC15}
\end{equation}
with $\overline{\overline{A_p}} =0.831$ .\\

20\% absorption
\begin{equation}
\Mavav(_{6}C^{12}) = \overline{\overline{A_p}}\left(\begin{array}{ccc}
{0.829}& {0.134} & {0.037} \\
{0.134}& {0.731} & {0.134} \\
{0.037}& {0.134} & {0.829}
\end{array}\right)
\label{eq:MC20}
\end{equation}
with $\overline{\overline{A_p}} =0.782$ .\\

25\% absorption
\begin{equation}
\Mavav(_{6}C^{12}) = \overline{\overline{A_p}}\left(\begin{array}{ccc}
{0.840}& {0.127} & {0.032} \\
{0.127}& {0.745} & {0.127} \\
{0.032}& {0.127} & {0.840}
\end{array}\right)
\label{eq:MC25}
\end{equation}
with $\overline{\overline{A_p}} =0.734$ .\\

\noindent\underline{Oxygen:} \\

15\% absorption
\begin{equation}
\Mavav(_{8}O^{16}) = \overline{\overline{A_p}}\left(\begin{array}{ccc}
{0.771} & {0.167} & {0.062} \\
{0.167} & {0.665} & {0.167} \\
{0.062} & {0.167} & {0.771}
\end{array}\right)
\label{eq:MO15}
\end{equation}
with $\overline{\overline{A_p}} =0.833$ .\\

20\% absorption
\begin{equation}
\Mavav(_{8}O^{16}) = \overline{\overline{A_p}}\left(\begin{array}{ccc}
{0.783} & {0.161} & {0.056} \\
{0.161} & {0.679} & {0.161} \\
{0.056} & {0.161} & {0.783}
\end{array}\right)
\label{eq:MO20}
\end{equation}
with $\overline{\overline{A_p}} =0.784$ .\\

25\% absorption
\begin{equation}
\Mavav(_{8}O^{16}) = \overline{\overline{A_p}}\left(\begin{array}{ccc}
{0.797} & {0.153} & {0.050} \\
{0.153} & {0.693} & {0.153} \\
{0.050} & {0.153} & {0.797}
\end{array}\right)
\label{eq:MO25}
\end{equation}
with $\overline{\overline{A_p}} =0.735$ .\\

30\% absorption
\begin{equation}
\Mavav(_{8}O^{16}) = \overline{\overline{A_p}}\left(\begin{array}{ccc}
{0.810} & {0.146} & {0.044} \\
{0.146} & {0.709} & {0.146} \\
{0.044} & {0.146} & {0.810}
\end{array}\right)
\label{eq:MO30}
\end{equation}
with $\overline{\overline{A_p}} =0.687$ .

\section{Forward- and backward charge exchange matrices}
\label{app:M3}
\noindent\underline{Carbon:} \\
\noindent 15\% absorption
\begin{equation}
\Mavav_+(_{6}C^{12}) = \overline{\overline{A_{p+}}}\left(\begin{array}{ccc}
{0.870} & {0.100} & {0.029} \\
{0.100} & {0.799} & {0.100} \\
{0.029} & {0.100} & {0.870}
\end{array}\right),
\Mavav_-(_{6}C^{12})=\overline{\overline{A_{p-}}}\left(\begin{array}{ccc}
{0.675} & {0.251}& {0.074} \\
{0.251} & {0.498} & {0.251} \\
{0.074} & {0.251} & {0.675}
\end{array}\right)
\label{eq:MC15m}
\end{equation}
with $\overline{\overline{A_{p+}}} =0.606$
and $\overline{\overline{A_{p-}}} =0.225$.\\

\noindent 20\% absorption
\begin{equation}
\Mavav_+(_{6}C^{12}) = \overline{\overline{A_{p+}}}\left(\begin{array}{ccc}
{0.880} & {0.094} & {0.026} \\
{0.094} & {0.811} & {0.094} \\
{0.026} & {0.094} & {0.880}
\end{array}\right),
\Mavav_-(_{6}C^{12})=\overline{\overline{A_{p-}}}\left(\begin{array}{ccc}
{0.685} & {0.247}& {0.068} \\
{0.247} & {0.505} & {0.247} \\
{0.068} & {0.247} & {0.685}
\end{array}\right)
\label{eq:MC20m}
\end{equation}
with $\overline{\overline{A_{p+}}} =0.578$
and $\overline{\overline{A_{p-}}} =0.204$.\\

\noindent 25\% absorption
\begin{equation}
\Mavav_+(_{6}C^{12}) = \overline{\overline{A_{p+}}}\left(\begin{array}{ccc}
{0.889} & {0.088} & {0.022} \\
{0.088} & {0.823} & {0.088} \\
{0.022} & {0.088} & {0.889}
\end{array}\right),
\Mavav_-(_{6}C^{12})=\overline{\overline{A_{p-}}}\left(\begin{array}{ccc}
{0.695} & {0.243}& {0.062} \\
{0.243} & {0.513} & {0.243} \\
{0.062} & {0.243} & {0.695}
\end{array}\right)
\label{eq:MC25m}
\end{equation}
with $\overline{\overline{A_{p+}}} = 0.549$
and $\overline{\overline{A_{p-}}} = 0.184$.\\

\noindent\underline{Oxygen:} \\

\noindent 15\% absorption
\begin{equation}
\Mavav_+(_{8}O^{16}) = \overline{\overline{A_{p+}}}\left(\begin{array}{ccc}
{0.829} & {0.125} & {0.046} \\
{0.125} & {0.750} & {0.125} \\
{0.046} & {0.125} & {0.829}
\end{array}\right),
\Mavav_-(_{8}O^{16})=\overline{\overline{A_{p-}}}\left(\begin{array}{ccc}
{0.635} & {0.265}& {0.100} \\
{0.265} & {0.470} & {0.265} \\
{0.100} & {0.265} & {0.635}
\end{array}\right)
\label{eq:MO15m}
\end{equation}
with $\overline{\overline{A_{p+}}} =0.581$
and $\overline{\overline{A_{p-}}} =0.252$.\\

\noindent 20\% absorption
\begin{equation}
\Mavav_+(_{8}O^{16}) = \overline{\overline{A_{p+}}}\left(\begin{array}{ccc}
{0.840} & {0.119} & {0.041} \\
{0.119} & {0.762} & {0.119} \\
{0.041} & {0.119} & {0.840}
\end{array}\right),
\Mavav_-(_{8}O^{16})=\overline{\overline{A_{p-}}}\left(\begin{array}{ccc}
{0.646} & {0.262}& {0.092} \\
{0.262} & {0.477} & {0.262} \\
{0.092} & {0.262} & {0.646}
\end{array}\right)
\label{eq:MO20m}
\end{equation}
with $\overline{\overline{A_{p+}}} =0.554$
and $\overline{\overline{A_{p-}}} =0.23$.\\

\noindent 25\% absorption
\begin{equation}
\Mavav_+(_{8}O^{16}) = \overline{\overline{A_{p+}}}\left(\begin{array}{ccc}
{0.852} & {0.112}& {0.036} \\
{0.112} & {0.776} & {0.112} \\
{0.036} & {0.112} & {0.852}
\end{array}\right),
\Mavav_-(_{8}O^{16})=\overline{\overline{A_{p-}}}\left(\begin{array}{ccc}
{0.657} & {0.258} & {0.085} \\
{0.258} & {0.485} & {0.257} \\
{0.085} & {0.258} & {0.657}
\end{array}\right)
\label{eq:MO25m}
\end{equation}
with $\overline{\overline{A_{p+}}} =0.527$
and $\overline{\overline{A_{p-}}} =0.208$.\\

\noindent 30\% absorption
\begin{equation}
\Mavav_+(_{8}O^{16}) = \overline{\overline{A_{p+}}}\left(\begin{array}{ccc}
{0.863} & {0.105} & {0.031} \\
{0.105} & {0.789} & {0.105} \\
{0.031} & {0.105} & {0.863}
\end{array}\right),
\Mavav_-(_{8}O^{16})=\overline{\overline{A_{p-}}}\left(\begin{array}{ccc}
{0.669} & {0.253} & {0.078} \\
{0.253} & {0.493} & {0.253} \\
{0.078} & {0.253} & {0.669}
\end{array}\right)
\label{eq:MO30m}
\end{equation}
with $\overline{\overline{A_{p+}}} =0.499$
and $\overline{\overline{A_{p-}}} =0.187$.
\end{appendix}

%\pagebreak

%\bibliographystyle{/afs/desy.de/user/s/schien/Bibliography/test_eprint}
%\bibliographystyle{/afs/desy.de/user/s/schien/Bibliography/myJHEP}
%\bibliography{/afs/desy.de/user/s/schien/Bibliography/electron}

%\bibliographystyle{/theo/yu/Bibliography/test_eprint}
%\bibliography{/theo/yu/Bibliography/electron}

%\bibliographystyle{/theo/schien/Bibliography/test_eprint}
%\bibliography{/theo/schien/Bibliography/electron}

%\bibliographystyle{/home/schien/Bibliography/test_eprint}
%\bibliography{/home/schien/Bibliography/electron}

%\bibliographystyle{/home/zylon/yu/PSY/test_eprint}
%\bibliography{/home/zylon/yu/PSY/electron}

\end{document}